\documentclass[journal,12pt,onecolumn,draftclsnofoot]{IEEEtran}

\usepackage{amsmath,amsfonts,amssymb,amsthm}
\usepackage{algorithm}
\usepackage{algorithmic}
\usepackage{array}
\usepackage[caption=false,font=normalsize,labelfont=sf,textfont=sf]{subfig}
\usepackage{textcomp}
\usepackage{dsfont}
\usepackage{stfloats}
\usepackage{url}
\usepackage{graphicx}
\usepackage{cite}
\usepackage{xcolor}
\usepackage{comment}
\usepackage{float}
\usepackage{multirow}
\usepackage{makecell}
\usepackage{bbm}
\usepackage{hyperref}
\newtheorem{theorem}{Theorem}

\newtheorem{proposition}{Proposition}
\newtheorem{corollary}{Corollary}
\theoremstyle{definition}
\newtheorem{definition}{Definition}
\newtheorem{remark}{Remark}

\DeclareMathOperator*{\argmax}{arg\,max}
\DeclareMathOperator*{\argmin}{arg\,min}
\newcommand{\R}{\mathbb{R}}
\newcommand{\E}{\mathbb{E}}

\newcommand{\vx}{\mathbf{x}}
\newcommand{\vy}{\mathbf{y}}
\newcommand{\vP}{\mathbf{P}}
\newcommand{\vQ}{\mathbf{Q}}
\newcommand{\vY}{\mathbf{Y}}
\newcommand{\vz}{\mathbf{z}}
\newcommand{\vd}{\mathbf{d}}
\newcommand{\vA}{\mathbf{A}}
\newcommand{\vRmat}{\mathbf{R}}
\newcommand{\vnu}{\boldsymbol{\nu}}
\newcommand{\calX}{\mathcal{X}}
\newcommand{\calY}{\mathcal{Y}}
\newcommand{\calV}{\mathcal{V}}
\newcommand{\calF}{\mathcal{F}}
\newcommand{\calS}{\mathcal{S}}

\begin{document}

\title{Reference-Dependent Utility Optimization for Goal-Oriented Communication Systems}

\author{Symeon~Vaidanis,~Photios~A.~Stavrou~and~Marios~Kountouris
\thanks{S. Vaidanis and P. A. Stavrou are with the Communication Systems Department of EURECOM, Sophia Antipolis, 06410 France, and M. Kountouris is with the Communication Systems Department of EURECOM, Sophia Antipolis, 06410 France and Department of Computer Science and Artificial Intelligence of University of Granada, C/ Periodista Daniel Saucedo Aranda, Granada, 18071 Spain. E-mail: symeon.vaidanis@eurecom.fr, fotios.stavrou@eurecom.fr, marios.kountouris@eurecom.fr, mariosk@ugr.es.}}

\maketitle

\begin{abstract}
Traditional resource allocation in communication systems usually optimizes operational performance metrics, such as rate, latency, energy efficiency, reliability, or weighted combinations thereof. Motivated by goal-oriented semantic communication, we develop a generalized allocation framework in which operational metrics are transformed through reference-dependent user utilities and subsequently evaluated by an orchestrator through expected utility, cumulative prospect-theoretic, or related criteria. The formulation separates the operational reward, the user-side value of that reward, and the system-level decision criterion under uncertainty. We characterize communication utilities that encode quality-of-service (QoS)/service reference levels, deficit sensitivity, post-reference saturation, and loss aversion-like behavior. Weighted sum-reward maximization, proportional fairness, $\alpha$-fairness, and stochastic expected utility maximization arise as exact special cases under appropriate specializations, while soft QoS-aware allocation is obtained through a deficit penalty generalization. For deterministic orthogonal systems, we derive utility-weighted marginal-reward KKT conditions that characterize the global solution when the composite user objectives are concave. For nonorthogonal and stochastic systems, we develop successive convex approximation (SCA) procedures and establish stationary point guarantees under explicit smooth or composite nonsmooth assumptions, including a branch-stationarity result for cumulative prospect theory (CPT) under eventual nondegenerate ordering identification. For globally solved penalized problems, we further derive an $O(1/\lambda)$ bound on the excess aggregate soft-QoS deficit, and on the deficit itself whenever a QoS-feasible allocation exists. We also obtain an $O(1/\epsilon^2)$ outer iteration bound for a smooth fixed-step SCA best-response residual. Numerical examples illustrate how reference points, marginal utility slopes, channel quality, and resource budgets shape the active user set and the resulting allocation policy.
\end{abstract}



\section{Introduction}
Resource allocation is a core component of communication system design. In conventional formulations, the designer directly optimizes operational performance metrics such as weighted sum rate, proportional fairness, delay, energy efficiency, or outage probability. These formulations are closely related to network utility maximization (NUM), in which resource allocation is interpreted through utility maximization, shadow prices, and proportional-fair operating points \cite{Kelly1998_NUM}. The $\alpha$-fair family further generalizes proportional fairness and includes limiting connections to max--min fairness \cite{MoWalrand2000}. The principles of utility maximization have also been central in wireless power control and rate allocation problems \cite{ChiangPowerControlGP}. This direct approach is mathematically convenient, but it does not explicitly distinguish an operational metric from the value that this metric provides to a user or application. This distinction becomes important in human-centric, semantic, goal-oriented, and service-aware networks, where an improvement in an operational metric may be highly valuable below a target operating point but provides only marginal value once the corresponding service requirement has been met~\cite{kountouris2021semantics,stavrou:2023, CPT_Alouini}.
Recent surveys further identify goal-oriented semantic communication as a key direction for 6G systems and highlight the need for context- and task-dependent performance criteria \cite{Getu2024GoalOrientedSurvey}.

Decision theory provides a complementary layer for systems operating under uncertainty. Expected Utility Theory (EUT) evaluates random prospects using objective probabilities \cite{EUT}, whereas Prospect Theory (PT) and Cumulative Prospect Theory (CPT) introduce reference dependence, loss aversion, and probability weighting to model behavioral decision making \cite{Kahneman_Tversky_1979,Kahneman_Tversky_1992}. These ideas have been applied to several communication problems, including human-centric communications \cite{CPT_Alouini}, random-access games \cite{CPT_Random_Access,CPT_Protocols}, pricing \cite{CPT_NOMA}, UAV-assisted networks \cite{CPT_UAV}, cognitive radio \cite{CPT_Cognitive}, anti-jamming communications \cite{CPT_Anti_Jamming}, and integrated sensing and communications (ISAC) \cite{CPT_ISAC}. Existing work typically embeds PT/CPT into a specific game or pricing mechanism \cite{CPT_Game_Theory}, while utility-proportional fairness formulations demonstrate how application-level utility can alter wireless resource allocation \cite{TychogiorgosUPF2012}. To the best of our knowledge, we are not aware of a unified resource allocation formulation that explicitly separates reference-dependent user-side valuation, system-level aggregation, and orchestrator-level decision-making under uncertainty. Importantly, these layers play distinct roles: the user-side utility represents the service-dependent value of an operational metric, whereas the orchestrator-level value functional determines how deterministic or uncertain system outcomes are evaluated. This distinction is particularly relevant to goal-oriented semantic communication, where the usefulness of information depends on the task, context, and effectiveness in achieving a prescribed goal.

In this work, each user obtains an operational metric $R_i$, such as rate, a delay-reliability score, a semantic-effectiveness score, or a task-completion metric. This metric is transformed through a user-side utility $u_i$, whose reference point represents a service target or QoS operating level. A system-level collection function combines these user values, and an orchestrator evaluates the resulting deterministic or random outcome using EUT, PT, CPT, or another value functional. This separation clarifies what is optimized, whose valuation is modeled, and how uncertainty is handled. In its general nonorthogonal or stochastic form, the resulting optimization problem can be nonconvex and nonsmooth, motivating algorithms that exploit the particular structure of the communication-allocation problem rather than treating it as generic black-box nonsmooth optimization \cite{JordanNonsmoothNonconvex2023}.

The main contributions are as follows.
\begin{itemize}

    \item \textit{Unified reference-dependent allocation framework}: We formulate a generalized resource allocation framework that explicitly separates the operational reward $R_i$, the user-side utility $u_i$, the system-level collection function $f$, and the orchestrator value functional $\calV$. This separation accommodates deterministic and stochastic allocation problems and allows service- and goal-dependent valuations to be modeled independently of the underlying physical metric.
    
    \item \textit{Reference-dependent utilities and classical special cases}: We characterize communication-oriented utilities with monotonicity, deficit sensitivity below a reference point, post-reference saturation, and loss aversion-like asymmetry. Under appropriate choices of utility, aggregation, and value functions, the framework recovers weighted sum-reward maximization, proportional fairness, $\alpha$-fairness, and stochastic expected-utility maximization as exact special cases, while soft QoS-aware allocation arises through an explicit deficit penalty. For globally solved penalized problems, we establish an $O(1/\lambda)$ bound on the excess aggregate soft-QoS deficit.
    
    \item \textit{Optimality structure for orthogonal allocation}: For deterministic orthogonal systems, we derive utility-weighted marginal-reward KKT conditions and show that they are sufficient for global optimality when the composite user objectives are concave. Under strict concavity, we obtain a unique generalized water-filling response characterized by the inverse marginal-utility map. We also derive a pointwise composite-concavity condition and specialize it to Shannon rewards, clarifying when a reference-dependent utility with a convex deficit branch can still yield a globally concave allocation problem.

    \item \textit{Algorithms and convergence for nonorthogonal and stochastic systems}: We develop a maximization-consistent SCA framework for deterministic nonorthogonal and scenario-based stochastic allocation. We establish stationary point guarantees for smooth and composite nonsmooth objectives, together with a CPT branch-stationarity result under eventual nondegenerate ordering identification. For the smooth fixed-step specialization, we derive an $O(1/\epsilon^2)$ outer-iteration bound for the best-response residual. We also provide dual projected-subgradient and ADMM-type implementations for the resulting convexified subproblems, while separating inner-solver convergence from that of the outer SCA procedure.

\end{itemize}

\section{Preliminaries in Decision and Utility Theory}
In this section, we review the key features of decision theory as established by von Neumann and Morgenstern through Expected Utility Theory (EUT) \cite{EUT}, together with the human-centric framework of Prospect Theory (PT) introduced by Kahneman and Tversky \cite{Kahneman_Tversky_1979} and its later extension, Cumulative Prospect Theory (CPT) \cite{Kahneman_Tversky_1992}.

\subsection{Decision-Theoretic Foundations: EUT and CPT}\label{subsec:axiomatic_EUT}

Let $L^{(a)}=\{(\omega_1,p_1^{(a)}),\ldots,(\omega_M,p_M^{(a)})\}$ denote lottery $a$ over the ordered outcome space $\calF=\{\omega_1,\ldots,\omega_M\}$, where $\omega_1\leq\cdots\leq\omega_M$ and $\sum_{m=1}^M p_m^{(a)}=1$. All lotteries are represented on this common finite outcome space, with zero probability allowed for outcomes outside a lottery's effective support. The lottery space is assumed to be closed under probabilistic mixtures, and $pL^{(1)}+(1-p)L^{(2)}$ denotes the corresponding reduced lottery. We write $L^{(1)}\succeq L^{(2)}$ for weak preference, $L^{(1)}\succ L^{(2)}$ for strict preference, and $L^{(1)}\sim L^{(2)}$ for indifference. The superscript indexes lotteries and is distinct from the user index introduced later. The EUT framework of von Neumann and Morgenstern is based on four standard rationality postulates \cite{EUT}:
\begin{enumerate}
    \item \emph{Completeness}: for any $L^{(1)},L^{(2)}$, either $L^{(1)}\succeq L^{(2)}$ or $L^{(2)}\succeq L^{(1)}$.
    \item \emph{Transitivity}: if $L^{(1)}\succeq L^{(2)}$ and $L^{(2)}\succeq L^{(3)}$, then $L^{(1)}\succeq L^{(3)}$.
    \item \emph{Continuity}: if $L^{(1)}\succeq L^{(2)}\succeq L^{(3)}$, then there exists $p\in[0,1]$ such that $pL^{(1)}+(1-p)L^{(3)}\sim L^{(2)}$.
    \item \emph{Independence}: for any $L^{(3)}$ and $p\in(0,1)$, $L^{(1)}\succeq L^{(2)}$ if and only if $pL^{(1)}+(1-p)L^{(3)}\succeq pL^{(2)}+(1-p)L^{(3)}$.
\end{enumerate}
Under these axioms, the preference relation admits an expected-utility representation as follows:
\begin{subequations}\label{eq:EUT_value}
\begin{align}
    L^{(1)}\succeq L^{(2)} &\Longleftrightarrow \calV_{\rm EUT}(L^{(1)})\geq \calV_{\rm EUT}(L^{(2)}),\\
    \calV_{\rm EUT}(L^{(a)})&=\sum_{m=1}^{M}p_m^{(a)}U(\omega_m),
\end{align}
\end{subequations}
where $U$ is unique up to a positive affine transformation. Risk neutrality corresponds to an affine Bernoulli utility, in which case lottery ranking is equivalent, up to a positive affine transformation, to ranking by expected outcome. A risk-neutral decision maker is recovered as the special case in which the utility over outcomes is affine, so that the value of a lottery reduces, up to an affine transformation, to the expected outcome. Nonlinear utility functions instead encode risk-sensitive preferences. A general monotone transformation preserves ordinal preferences but, in general, does not preserve the expected-utility form.

PT and CPT relax the linear treatment of probabilities in EUT. In PT, each probability mass is distorted individually, whereas in CPT \cite{WakkerTve93}, the cumulative probabilities of ordered gains and losses are distorted. Because $\omega_1\leq\cdots\leq\omega_M$, let $\mathcal I^-=\{m:\omega_m<\omega_0\}$ denote the loss indices in increasing outcome order and $\mathcal I^+=\{m:\omega_m\geq\omega_0\}$ the gain indices in increasing outcome order. Outcomes equal to the reference point are assigned to $\mathcal I^+$ by convention; since $U(\omega_0)=0$, this assignment does not affect the CPT value. The CPT value is
\begin{equation}\label{eq:CPT_value}
    \calV_{\rm CPT}(L^{(a)})=\sum_{m\in\mathcal I^-} c_{m,-}^{(a)}U(\omega_m)+\sum_{m\in\mathcal I^+} c_{m,+}^{(a)}U(\omega_m),
\end{equation}
where
\begin{subequations}\label{eq:CPT_weights}
\begin{align}
    c_{m,-}^{(a)} &= W^-\!\left(\sum_{j\in\mathcal I^-:j\le m}p_j^{(a)}\right)-W^-\!\left(\sum_{j\in\mathcal I^-:j<m}p_j^{(a)}\right),\\
    c_{m,+}^{(a)} &= W^+\!\left(\sum_{j\in\mathcal I^+:j\ge m}p_j^{(a)}\right)-W^+\!\left(\sum_{j\in\mathcal I^+:j>m}p_j^{(a)}\right).
\end{align}
\end{subequations}
The separation between gains and losses avoids the index ambiguity and possible double counting that arise when a single cumulative expression is used around the reference point. The PT weights are recovered by replacing the cumulative decision weights $c_{m,\pm}^{(a)}$ with probability-mass weights $W^\pm(p_m^{(a)})$.

\subsection{Reference-Dependent Utility Functions}\label{subsec:utility_function}
The role of a utility or value function depends on the modeling layer. In decision theory, $U$ maps outcomes into subjective value and, together with the probability or decision-weighting rule, induces preferences over random prospects. In utility-theoretic resource allocation, $u_i$ maps an operational metric $R_i$ to the value obtained by user $i$. This paper uses both layers: $u_i$ models the user-side utilization of communication resources, whereas $U_{\rm orch}$ models the orchestrator's valuation of outcomes under uncertainty.

A CPT value function is reference dependent. Given a reference point $\omega_0$, we normalize $U(\omega_0)=0$; outcomes above $\omega_0$ are gains, whereas outcomes below $\omega_0$ are losses. The classical S-shaped CPT value function is convex over losses and concave over gains. For a symmetric fair binary gamble around $\omega_0$, strict loss aversion implies
\begin{equation}\label{eq:sym_bet_aversion}
U(\omega_0+\delta)-U(\omega_0) < U(\omega_0)-U(\omega_0-\delta),\quad \delta>0.
\end{equation}
When $U$ is differentiable on both sides of the reference point, a sufficient differential condition for increasing aversion to symmetric bets is
\begin{equation}\label{eq:inc_sym_bet_aversion}
U'(\omega_0+\delta)<U'(\omega_0-\delta),\quad \delta>0.
\end{equation}
For nonsymmetric fair binary gambles with $\delta_1,\delta_2>0$ and outcomes $\omega_0+\delta_2$ and $\omega_0-\delta_1$, the fair probabilities of the gain and loss outcomes are $\delta_1/(\delta_1+\delta_2)$ and $\delta_2/(\delta_1+\delta_2)$, respectively. Neilson's weak loss aversion can therefore be stated as
\begin{equation}\label{eq:neilson_weak}
\frac{\delta_1}{\delta_1+\delta_2}U(\omega_0+\delta_2)+\frac{\delta_2}{\delta_1+\delta_2}U(\omega_0-\delta_1)<U(\omega_0),
\end{equation}
or, equivalently, the average loss-side slope dominates the average gain-side slope,
\begin{equation}\label{eq:neilson_weak_slope}
\frac{U(\omega_0+\delta_2)-U(\omega_0)}{\delta_2}
<
\frac{U(\omega_0)-U(\omega_0-\delta_1)}{\delta_1}.
\end{equation}
Neilson's strong loss aversion imposes the corresponding slope-dominance condition for nested gambles,
\begin{equation}\label{eq:neilson_strong}
\frac{U(\omega_0+\delta_4)-U(\omega_0+\delta_2)}{\delta_4-\delta_2}
<
\frac{U(\omega_0-\delta_3)-U(\omega_0-\delta_1)}{\delta_1-\delta_3},
\end{equation}
for $0\leq\delta_3<\delta_1$ and $0\leq\delta_2<\delta_4$ \cite{Neilson_2002_A, vaidanis2025Theoretical_Part_1_Neilson_extension_KW}.
These conditions concern loss aversion in the decision-theoretic sense. In the communication-allocation model developed later, analogous gain/loss asymmetries may instead be used to describe differential sensitivity to service surplus and deficit around a reference operating level.

\subsection{Probability Weighting Functions}\label{subsec:PWF}
A probability weighting function (PWF) is a continuous, strictly increasing map $W^{\pm}:[0,1]\rightarrow[0,1]$ satisfying $W^{\pm}(0)=0$ and $W^{\pm}(1)=1$. In PT, probability masses are transformed directly, whereas CPT applies weighting to cumulative probabilities. With a nondecreasing value function and monotone cumulative weighting, the CPT representation preserves first-order stochastic dominance under the standard conditions \cite{Kahneman_Tversky_1992,Wakker_Book}. The gain and loss weighting functions $W^{+}$ and $W^{-}$ may be chosen separately and may use different parameter values. The Tversky-Kahneman PWF is
\begin{equation}
    W(p)=\frac{p^{\delta}}{\left(p^{\delta}+(1-p)^{\delta}\right)^{1/\delta}},\quad 0<\delta\leq 1,
\end{equation}
where the parameter must in addition be chosen so that this parametric form is strictly increasing on $[0,1]$, as required by the PWF definition above. The two-parameter Prelec PWF is
\begin{equation}\label{eq:Prelec_PWF}
    W(p)=\exp\!\left[-\gamma\left(-\ln p\right)^{\theta}\right],\quad \gamma>0,\;\theta>0, \quad 0<p\leq1.
\end{equation}
For $p=0$, $W(0)=0$ is defined by continuity. For $0<\theta<1$, the Prelec function is inverse S-shaped; $\gamma$ controls the elevation and the crossing point relative to $W(p)=p$ \cite{Prelec}. For $\theta=1$, the Prelec form reduces to $W(p)=p^{\gamma}$.

\subsection{Extended K\"obberling-Wakker Utility}\label{subsec:extension_KW}
An extension of the K\"obberling-Wakker utility \cite{Kobberling_Wakker}, used in the communication-allocation setting of \cite{vaidanis2025ICC,vaidanis2025Theoretical_Part_1_Neilson_extension_KW}, is defined as follows:
\begin{equation}\label{eq:Extension_KW}
    u(\omega)\triangleq
        \begin{cases}
            \lambda_1 \dfrac{1-\exp\!\left(\frac{\alpha}{\gamma_1}g_1(\omega,\omega_0)\right)}{\alpha}, & \omega \ge \omega_0,\\[6pt]
            \lambda_2 \dfrac{1-\exp\!\left(\frac{\beta}{\gamma_2}g_2(\omega,\omega_0)\right)}{\beta}+\lambda_3(\omega-\omega_0), & \omega < \omega_0.
        \end{cases}
\end{equation}
The parameters $\gamma_1$ and $\gamma_2$ tune local curvature, while $\lambda_1$, $\lambda_2$, and $\lambda_3$ control marginal slopes and relative gain/loss valuation. The branch functions $g_1$ and $g_2$ must be specified on the gain and loss domains, respectively, and satisfy $g_1(\omega_0,\omega_0)=g_2(\omega_0,\omega_0)=0$ to preserve the normalization $u(\omega_0)=0$. Their derivatives and parameter signs must be chosen so that the resulting utility is nondecreasing and has the intended curvature on each branch. In communication systems, $\omega_0$ is interpreted as a target operating point, typically a QoS or service reference level. Equation~\eqref{eq:Extension_KW} is therefore a parametric template rather than an assertion that every parameter choice has the desired shape. Any concrete use of this template must specify $g_1$, $g_2$, and the parameter vector and verify the required monotonicity, curvature, and regularity conditions on the operating interval.

The preceding concepts provide the utility and probability weighting tools used below. We next embed them into a resource allocation model that separates physical rewards, user-side utility transformations, and system-level decision criteria.

\section{Generalized Resource Allocation Framework}\label{sec:framework}
This section introduces the proposed generalized resource allocation framework. The objective is to decouple the generation of operational rewards from their user-side utility transformation and the corresponding system-level decision-theoretic evaluation. This separation provides a unified model for deterministic and stochastic allocation problems in which users may have heterogeneous service requirements, reference points, and sensitivities to surplus or deficit relative to their target operating levels.

\subsection{System Model}\label{subsec:system_model}
Consider $N$ users served by a common orchestrator. User $i$ receives resource $x_i$, and the vector of resources is $\vx=[x_1,\ldots,x_N]^T\in\calX$. The operational metric of user $i$ is $R_i(\vx,\vY)$, where $\vY$ denotes the random system state, such as channel gains, traffic arrivals, interference levels, or task states. In an orthogonal system, $R_i(\vx,\vY)=R_i(x_i,Y_i)$; in a nonorthogonal system, $R_i$ depends on the entire vector $\vx$ and possibly on the full state $\vY$.

The resource vector may represent transmit powers, bandwidth shares, computing cycles, access probabilities, semantic compression levels, or other controllable allocation variables. The feasible set $\calX$ therefore represents the network architecture and the associated resource constraints. Per-user or uplink-type constraints can be represented by
\begin{equation}\label{eq:per_user_resource_set}
    \calX_{\rm per}=\{\vx:\;0\leq x_i\leq x_i^{\max},\; i=1,\ldots,N\},
\end{equation}
whereas shared downlink or orchestrator-side budget constraints can be represented by
\begin{equation}\label{eq:shared_resource_set}
    \calX_{\rm sh}=\{\vx:\;x_i\geq0,\;\sum_{i=1}^{N}x_i\leq X_{\rm tot}\}.
\end{equation}
The two forms can also be combined when both individual and aggregate constraints are present. The orthogonal/nonorthogonal distinction concerns coupling in the reward map: orthogonal models satisfy $R_i(\vx,\vY)=R_i(x_i,Y_i)$, whereas nonorthogonal models may include interference, shared queues, or coupled computation so that $R_i$ depends on multiple allocation variables. Statistical dependence among the state components $Y_i$ is a separate modeling issue and, by itself, does not make the allocation model nonorthogonal.

\begin{figure}[t]
    \centering
    \includegraphics[width=0.7\linewidth]{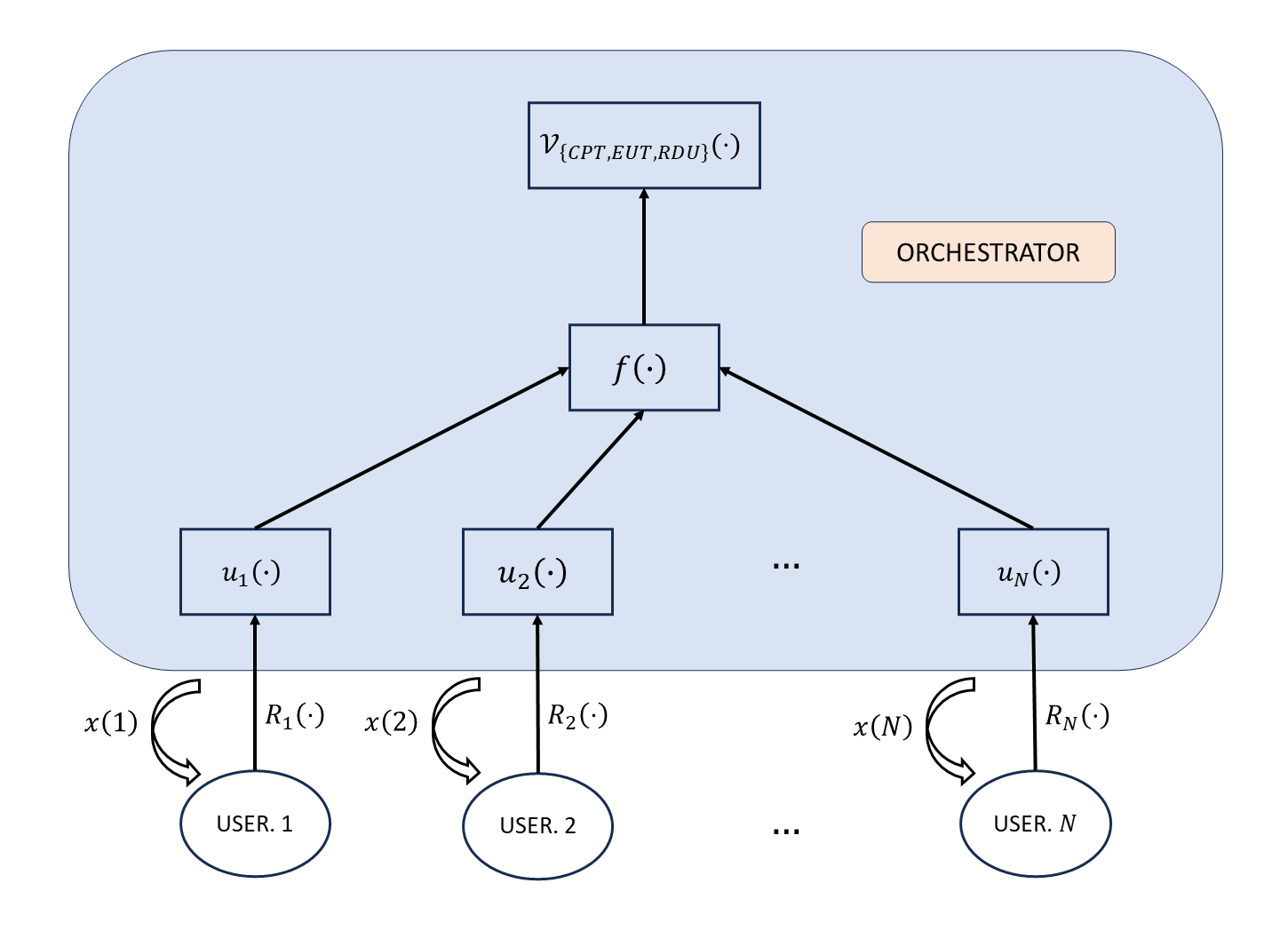}
    \caption{Proposed generalized resource allocation framework. The physical system first generates operational rewards, which are transformed through user-side utility functions and then evaluated by an orchestrator using a deterministic or decision-theoretic value functional.}
    \label{fig:system_model}
\end{figure}

Fig.~\ref{fig:system_model} illustrates the proposed separation among the operational reward, the user-side utility transformation, and the orchestrator-level decision criterion.

The proposed framework applies the utility map $u_i$ to the operational metric and combines the resulting user-side utilities through a collection function $f$:
\begin{equation}\label{eq:collected_reward}
G(\vx,\vY)=f\left(u_1(R_1(\vx,\vY)),\ldots,u_N(R_N(\vx,\vY))\right).
\end{equation}
The orchestrator evaluates $G(\vx,\vY)$ using a value functional $\calV_{\rm orch}$. For deterministic systems, choosing the identity orchestrator value, or more generally any strictly increasing scalar transformation of $G(\vx)$, yields the same maximizers as directly maximizing $G(\vx)$. For stochastic systems, $\calV_{\rm orch}$ may be an expected utility, a CPT value, or a risk-sensitive criterion. A generic allocation problem is therefore given by
\begin{equation}\label{eq:generic_problem}
\begin{aligned}
    \max_{\vx\in\calX}\quad & \calV_{\rm orch}\!\left(G(\vx,\vY)\right)\\
    \text{s.t.}\quad & \text{optional hard QoS or service constraints}.
\end{aligned}
\end{equation}
The hard constraints remain distinct from the utility functions. Utilities can encourage service-aware behavior, but do not, by themselves, guarantee deterministic QoS satisfaction. Hard deterministic guarantees require explicit constraints; penalty formulations provide soft enforcement and yield QoS guarantees only under additional assumptions, as discussed later.

This separation is important because the utility map $u_i$ is not a replacement for feasibility requirements. It describes how user $i$ values a deficit or surplus of the operational metric relative to its service target. Hard QoS constraints, admission constraints, and outage constraints should remain explicit when deterministic guarantees are required. Conversely, when the goal is to trade service deficits against surplus utility, the same reference point can be used inside a soft penalty or value function.

When the system state is random, the uncertainty should be modeled using joint state realizations rather than independent per-user samples, unless independence is explicitly justified. A finite scenario representation of the random state $\vY$ is given by
\begin{equation}\label{eq:joint_scenario_set_framework}
\widehat{\calY}=\{(\vy^{(k)},p_k)\}_{k=1}^{K},\qquad p_k\geq0,\quad \sum_{k=1}^{K}p_k=1,
\end{equation}
where $\vy^{(k)}$ denotes the $k$-th joint realization of $\vY$, including channels, traffic, tasks, or semantic events. Each allocation $\vx$ induces the system-level prospect
\begin{equation}\label{eq:allocation_induced_prospect}
    L_{\rm orch}(\vx)=\{(G_k(\vx),p_k)\}_{k=1}^{K},\quad
    G_k(\vx)=G(\vx,\vy^{(k)}),
\end{equation}
which is evaluated by $\calV_{\rm orch}$. User-side stochastic service requirements can be written using potentially distinct value functionals, for example,
\begin{equation}\label{eq:user_side_stochastic_qos}
    \calV_i\!\left(\{(u_i(R_i(\vx,\vy^{(k)})),p_k)\}_{k=1}^{K}\right)\geq \eta_i,\quad i=1,\ldots,N.
\end{equation}
Thus, the orchestrator's uncertainty preference and the users' service-evaluation rules need not coincide.

For a downlink power allocation example, $x_i=P_i$, $\sum_iP_i\leq P_{\rm tot}$, and $P_i\geq0$. For an interference-limited model, a standard rate expression is
\begin{equation}\label{eq:interference_rate}
    R_i(\vP,\vY)=\log_2\!\left(1+\frac{g_{ii}P_i}{\sum_{j\neq i}g_{ij}P_j+N_0}\right),
\end{equation}
where $g_{ij}=g_{ij}(\vY)$ is the channel/interference gain from transmitter or stream $j$ to receiver $i$. For power-domain downlink NOMA, the same framework can be used, but the decoding order, SIC constraints, and possible residual interference terms must be explicitly specified.

The classical physical-layer allocation model is recovered by choosing $u_i(r)=r$, a weighted-sum collection function, and a risk-neutral expected-value criterion at the orchestrator layer. Therefore, the proposed framework does not replace standard rate models; rather, it inserts an intermediate utility and decision-theoretic layer between the operational reward and the final system objective.

\subsection{Semantic-Aware Probability-Weighted Power Allocation}\label{subsec:semantic_cpt_power_allocation}
The framework also covers the semantic-aware probability-weighted allocation model studied in \cite{vaidanis2025SPAWC}. This formulation is PT/CPT-inspired: reference-dependent utilities and individually transformed event probabilities model subjective user-side valuation, but no outcome ranking or cumulative decision weights are used. Full rank-dependent CPT evaluation, in which scenario outcomes are ordered and cumulative decision weights are computed, is considered separately in Section~\ref{sec:stochastic}.

In goal-oriented semantic communication, the operational metric $R_i$ need not be a Shannon rate; it can be a semantic-effectiveness score, a task-completion probability, a fidelity metric, an information-importance score, or an SNR-driven surrogate for one of these quantities. Let $p_i$ denote the objective probability that user $i$ is active, that a semantic event is relevant, or that a goal-dependent information item is required. The user's subjective perception of this probability can be represented by a PWF $W_i(p_i)$. A semantic-aware power allocation instance is then given by
\begin{equation}\label{eq:semantic_cpt_power}
\begin{aligned}
    \max_{\vP}\quad & \sum_{i=1}^{N} W_i(p_i) u_i\!\left(S_i(P_i,h_i)\right)\\
    \text{s.t.}\quad & P_i\geq0,\quad i=1,\ldots,N,\\
    & \sum_{i=1}^{N}P_i\leq P_{\rm tot},
\end{aligned}
\end{equation}
where $S_i(P_i,h_i)$ is the semantic or physical effectiveness metric. For the SNR-driven specialization,
\begin{equation}\label{eq:snr_semantic_metric}
    S_i(P_i,h_i)=\frac{P_i|h_i|^2}{\sigma^2}.
\end{equation}
Equation~\eqref{eq:semantic_cpt_power} should be interpreted as a subjective information-significance maximization problem: $W_i(p_i)$ captures the perceived likelihood or importance of the event, while $u_i$ captures the reference-dependent value of the achieved operating level. This formulation is separable in orthogonal channels, but for general reference-dependent or CPT-inspired utility specifications the resulting objective may be nonconcave, nonsmooth, or both.

\subsection{Decision Theory Versus Coherent Risk Measures}\label{subsec:comparison_decision_theory_vs_coherent_risk_measure}
Decision-theoretic value functions and coherent risk measures, for instance Conditional Value at Risk (CVaR) \cite{RockafellarUryasev2000}, both evaluate uncertain quantities, but they are based on different axiomatic formulations. With the common sign convention that $\rho$ evaluates a random loss variable, a coherent risk measure satisfies monotonicity, translation invariance, positive homogeneity, and subadditivity \cite{coherent_risk_measures}:
\begin{align}
    X\leq Y &\Rightarrow \rho(X)\leq\rho(Y),\\
    \rho(X+c)&=\rho(X)+c,\quad c\in\R,\\
    \rho(\lambda X)&=\lambda\rho(X),\quad \lambda\geq0,\\
    \rho(X+Y)&\leq\rho(X)+\rho(Y).
\end{align}
If $X$ is a payoff rather than a loss, the monotonicity and translation signs must be adjusted accordingly. 

Coherent risk measures and decision-theoretic value functions answer different modeling questions. Decision-theoretic axioms specify how allocation-induced prospects, i.e., probability distributions over outcomes, are ranked according to a preference relation. In contrast, coherent-risk-measure axioms specify structural properties of a risk functional acting on a random loss variable; the distinction is therefore between prospect evaluation and loss-risk assessment, rather than between different event spaces. A coherent risk measure maps a loss random variable to a scalar risk score and is most appropriate when the engineering objective is to control downside exposure, such as tail loss, outage cost, or reliability violation. By contrast, EUT and CPT rank allocation-induced prospects through a utility-based preference relation; CPT additionally incorporates reference dependence and probability weighting. Therefore, in the present framework, EUT/CPT criteria are used when the objective is subjective valuation or behavioral perception of communication outcomes, whereas CVaR-type tools are more appropriate when a tail-risk guarantee or downside-risk constraint is required. The two approaches can also be combined, for example by including CVaR as a constraint, a penalty term, or an additional orchestrator-level risk criterion.

\subsection{Special Cases and Scope}
The proposed formulation includes several classical communication objectives as special cases.

\begin{proposition}[Exact special cases and soft-penalty generalizations]\label{thm:special_cases}
For the deterministic specialization of \eqref{eq:generic_problem}, let the collection function be linear, $f(z_1,\ldots,z_N)=\sum_{i=1}^{N}w_i z_i$, with $w_i\geq0$ and $\sum_iw_i>0$. Then items $1$-$3$ are exact classical special cases, item $4$ is a probability-weighted specialization, and item $5$ is a soft-penalty generalization. Item $6$ is an exact special case of the stochastic formulation:
\begin{enumerate}
    \item $u_i(r)=r$ yields weighted sum-reward maximization, and weighted sum-rate maximization when $R_i$ is a rate.
    \item $u_i(r)=\log r$ for $r>0$ yields proportional fairness. Ordinary proportional fairness is recovered when the positive weights are equal.
    \item $u_i(r)=(r^{1-\alpha}-1)/(1-\alpha)$ for $r>0$, $\alpha\geq0$, and $\alpha\neq1$, together with $u_i(r)=\log r$ for $\alpha=1$, yields the weighted normalized $\alpha$-fair family and is continuous at $\alpha=1$; the usual unweighted family is recovered for equal positive weights.
    \item With reference-dependent $u_i$ and weights $w_i=W_i(p_i)$, the formulation recovers the probability-weighted allocation in \eqref{eq:semantic_cpt_power}; this is not full rank-dependent CPT.  
    \item A reference-dependent utility with an explicit deficit penalty below $r_i^0$ yields a soft QoS-aware allocation; the limiting behavior is formalized in Theorem~\ref{thm:soft_qos}.
    \item If $G(\vx,\vY)$ is random and $\calV_{\rm orch}(G)=\E[U_{\rm orch}(G)]$, then \eqref{eq:generic_problem} reduces to stochastic expected-utility maximization.
\end{enumerate}
\end{proposition}
\begin{proof}
Let
\begin{equation*}
        J(\vx)\triangleq \sum_{i=1}^{N}w_i u_i(R_i(\vx)).
\end{equation*}
For $u_i(r)=r$,
\begin{equation*}
J(\vx)=\sum_{i=1}^{N}w_iR_i(\vx),
\end{equation*}
which is weighted sum-reward maximization and becomes weighted sum-rate maximization when $R_i$ is a rate. For $u_i(r)=\log r$,
\begin{equation*}
J(\vx)=\sum_{i=1}^{N}w_i\log R_i(\vx),
\end{equation*}
which is the weighted proportional-fair objective. For $\alpha\neq1$, the weighted normalized $\alpha$-fair objective is
\begin{equation*}
J_{\alpha}(\vx)=\sum_{i=1}^{N}w_i\frac{R_i(\vx)^{1-\alpha}-1}{1-\alpha},
\end{equation*}
and
\begin{equation*}
\lim_{\alpha\to1}\frac{r^{1-\alpha}-1}{1-\alpha}=\log r,
\end{equation*}
so the family is continuous at $\alpha=1$. Setting $w_i=W_i(p_i)$ and $R_i=S_i(P_i,h_i)$ gives exactly the objective in \eqref{eq:semantic_cpt_power}. For a deficit-penalized utility
\begin{equation*}
u_i(r)=s_i(r)-\lambda d_i\bigl((r_i^0-r)_+\bigr),
\end{equation*}
the aggregate objective becomes $S(\vx)-\lambda D(\vx)$, whose limiting behavior is established in Theorem~\ref{thm:soft_qos}. Finally, if $G(\vx,\vY)$ is random and $\calV_{\rm orch}(G)=\E[U_{\rm orch}(G)]$, then
\begin{equation*}
\calV_{\rm orch}(G(\vx,\vY)) =\int U_{\rm orch}(G(\vx,\vy))\,\mathrm dP_{\vY}(\vy),
\end{equation*}
which is precisely stochastic expected-utility maximization.
\end{proof}

Having shown that the proposed formulation contains several classical objectives as limiting or special cases, we now specify the reference-dependent utility functions that enable service-aware and goal-oriented allocation behavior.

\section{Communication-Oriented Utility Design}\label{sec:utility_design}
This section specializes the utility layer introduced in Section~\ref{sec:framework} by specifying communication-oriented reference-dependent utility functions that map operational rewards into service-aware user value.

A reference-dependent utility for resource allocation should be increasing in the operational metric, should assign high marginal value to improvements below the reference point, and may saturate above the reference point. Let $r_i^0$ denote the target operating level of user $i$, and normalize $u_i(r_i^0)=0$.

This utility is designed from a utilization perspective rather than as a substitute for the operational reward. Below $r_i^0$, the user operates in a deficit region, where additional reward is particularly valuable because the target service level has not yet been achieved. Above $r_i^0$, additional reward may still be useful, but its marginal value typically decreases because the service requirement has already been met. We refer to these two effects as deficit necessity and surplus saturation, respectively. At a differentiable reference point, the one-sided derivatives coincide. Loss-aversion-like asymmetry can nevertheless be represented smoothly through different gain- and loss-branch behavior away from the reference point. A kink instead models an immediate slope discontinuity, with a larger left derivative than right derivative.

For utilities that are differentiable on each side of the reference point, the preceding interpretation motivates the following branchwise shape conditions:
\begin{subequations}\label{eq:utility_shape_conditions}
\begin{align}
    u_i'(r)&\geq0,\qquad r\geq0,\ r\neq r_i^0,\label{eq:utility_monotone_condition}\\
    u_i''(r)&\geq0,\qquad r<r_i^0,\label{eq:utility_deficit_convexity}\\
    u_i''(r)&\leq0,\qquad r>r_i^0.\label{eq:utility_surplus_concavity}
\end{align}
\end{subequations}
The first condition states that a larger operational reward should not reduce user value. The second condition models increasing marginal value in the deficit region as the user approaches the target from below. The third condition models diminishing marginal value in the surplus region after the target has been reached.

\begin{definition}[Reference-dependent communication utility]
A function $u_i:\R_+\rightarrow\R$ is a reference-dependent communication utility with reference point $r_i^0$ if it is continuous, nondecreasing, satisfies $u_i(r_i^0)=0$, is convex for $r<r_i^0$, and is concave for $r>r_i^0$. We call the reference point \textit{slope-matched} if the finite one-sided derivatives satisfy $u_i'(r_i^{0-})=u_i'(r_i^{0+})$, and \textit{kinked} if $u_i'(r_i^{0-})>u_i'(r_i^{0+})$. Whenever a later SCA result invokes a ``smooth'' utility, we use the stronger optimization regularity assumption that $u_i$ is continuously differentiable on a neighborhood of the relevant operating interval and that $u_i'$ is locally Lipschitz there. Equality of the one-sided slopes alone does not imply this stronger property.
\end{definition}

Convexity below $r_i^0$ models increasing marginal necessity as the user approaches the target from below; concavity above $r_i^0$ models diminishing marginal value or saturation, after the target has been met.

\begin{remark}[Utility functions and QoS constraints]\label{rem:utility_not_qos}
The reference level $r_i^0$ should not be interpreted as a guaranteed QoS constraint unless the optimization problem includes an explicit constraint of the form $R_i(\vx)\geq r_i^0$ or a stochastic counterpart such as \eqref{eq:user_side_stochastic_qos}. The role of $u_i$ is to shape the allocation policy by rewarding surplus and penalizing deficit; however, it does not render infeasible service guarantees feasible. Likewise, a finite soft-penalty parameter does not in general enforce exact target satisfaction. Theorem~\ref{thm:soft_qos} establishes the corresponding asymptotic behavior under global optimization.
\end{remark}

Beyond monotonicity and curvature, loss aversion-like asymmetry can be quantified by comparing the gain and loss branches around the reference point. In a communication-utility interpretation, symmetric loss aversion-like asymmetry means that an equal deficit below $r_i^0$ has a larger absolute utility impact than an equal surplus above $r_i^0$, consistent with \eqref{eq:sym_bet_aversion}. A stronger local condition compares marginal utilities: increasing symmetric loss aversion requires the left-side marginal utility to dominate the right-side marginal utility in a neighborhood of $r_i^0$, consistent with \eqref{eq:inc_sym_bet_aversion}. These comparisons motivate kinked utilities with unequal one-sided slopes and smooth utilities whose gain and loss branches become asymmetric away from the reference point.

\begin{theorem}[Soft QoS limit]\label{thm:soft_qos}
    Let $\calX$ be nonempty and compact, and let $R_i$ be continuous. Define
    \begin{equation}
        F_{\lambda}(\vx)=S(\vx)-\lambda D(\vx),
    \end{equation}
    where $S$ is continuous and bounded on $\calX$, $D(\vx)=\sum_{i=1}^{N}w_i d_i((r_i^0-R_i(\vx))_+)$, $w_i\geq0$, and each $d_i$ is continuous, nonnegative, and satisfies $d_i(0)=0$. For every user with $w_i>0$, assume in addition that $d_i(t)>0$ for all $t>0$ (equivalently, $d_i(t)=0$ if and only if $t=0$ on $\R_+$). Let $\vx_{\lambda}\in\argmax_{\vx\in\calX}F_{\lambda}(\vx)$. For any sequence $\lambda_n\to\infty$ and any choice $\vx_{\lambda_n}\in\argmax_{\vx\in\calX}F_{\lambda_n}(\vx)$, every accumulation point of $\{\vx_{\lambda_n}\}$ minimizes $D$ over $\calX$. Moreover, every such accumulation point maximizes $S$ over the set of deficit-minimizing allocations, $\arg\min_{\vx\in\mathcal{X}} D(\vx)$. If there exists $\vx\in\calX$ such that $R_i(\vx)\geq r_i^0$ for all users with $w_i>0$, then every accumulation point satisfies the QoS targets of all positively weighted users. In this case, $D^\star=0$, and the accumulation points maximize $S$ over the zero-deficit set $\{\vx\in\mathcal{X}:D(\vx)=0\}$. If, in addition, $w_i>0$ for all $i$, this set coincides with the QoS-feasible set.
\end{theorem}
\begin{proof}
Compactness of $\calX$ and continuity of $D$ imply that
    \begin{equation*}
        D^\star\triangleq\min_{\vx\in\calX}D(\vx)
    \end{equation*}
is attained. Let $\vx^D\in\argmin_{\vx\in\calX}D(\vx)$ and define
    \begin{equation*}
        M_S\triangleq\max_{\vx\in\calX}|S(\vx)|<\infty.
    \end{equation*}
By optimality of $\vx_{\lambda_n}$,
    \begin{align*}
    S(\vx_{\lambda_n})-\lambda_nD(\vx_{\lambda_n})
    &\geq S(\vx^D)-\lambda_nD^\star,
    \end{align*}
    so that
    \begin{align}
        0\leq D(\vx_{\lambda_n})-D^\star
        &\leq \frac{S(\vx_{\lambda_n})-S(\vx^D)}{\lambda_n}
        \leq \frac{2M_S}{\lambda_n}.
        \label{eq:soft_qos_deficit_bound}
    \end{align}
Hence $D(\vx_{\lambda_n})\to D^\star$. Let $\vx_{\lambda_{n_j}}\to\bar\vx$ be any convergent subsequence, which exists by compactness. Continuity of $D$ and \eqref{eq:soft_qos_deficit_bound} yield
\begin{equation*}
D(\bar\vx)=\lim_{j\to\infty}D(\vx_{\lambda_{n_j}})=D^\star,
\end{equation*}
so every accumulation point minimizes $D$.

We next show that every such accumulation point also maximizes
$S$ over the deficit-minimizing set. Let $\vx^D\in\arg\min_{\vx\in\mathcal{X}}D(\vx)$ be arbitrary.
Optimality gives
\begin{equation*}
S(\vx_{\lambda_n})-\lambda_n D(\vx_{\lambda_n})
\geq
S(\vx^D)-\lambda_n D^\star.
\end{equation*}
Since $D(\vx_{\lambda_n})\geq D^\star$, it follows that
\begin{equation*}
S(\vx_{\lambda_n})-S(\vx^D)
\geq
\lambda_n\bigl(D(\vx_{\lambda_n})-D^\star\bigr)
\geq 0.
\end{equation*}
Passing to the convergent subsequence and using continuity of
$S$ gives
\begin{equation*}
S(\bar{\vx})\geq S(\vx^D),
\qquad
\forall\,
\vx^D\in
\arg\min_{\vx\in\mathcal{X}}D(\vx).
\end{equation*}
Thus every accumulation point maximizes $S$ over
$\arg\min_{\vx\in\mathcal{X}}D(\vx)$.

If there exists $\vx\in\mathcal{X}$ satisfying $R_i(\vx)\geq r_i^0$ for every user with $w_i>0$, then $D^\star=0$. Consequently, every accumulation point satisfies
$D(\bar{\vx})=0$. Since $w_i>0$ and
$d_i(t)>0$ for every $t>0$, each positively weighted summand
can vanish only if $(r_i^0-R_i(\bar{\vx}))_+=0$. Hence $R_i(\bar{\vx})\geq r_i^0$ for every user with $w_i>0$.

In this case,
$\arg\min_{\vx\in\mathcal{X}}D(\vx)
=
\{\vx\in\mathcal{X}:D(\vx)=0\}$,
so the preceding argument shows that every accumulation point
maximizes $S$ over the zero-deficit set. If $w_i>0$ for all
$i$, this set coincides with the QoS-feasible set.
\end{proof}

\begin{corollary}[Quantitative soft-QoS deficit bound]\label{cor:soft_qos_rate}
Under the assumptions of Theorem~\ref{thm:soft_qos}, let $M_S\triangleq\max_{\vx\in\calX}|S(\vx)|$. For every $\lambda>0$ and every global maximizer $\vx_\lambda$ of $F_\lambda$, we have
\begin{equation}\label{eq:soft_qos_rate_cor}
0\leq D(\vx_\lambda)-D^\star\leq\frac{2M_S}{\lambda}.
\end{equation}
If there exists $\vx\in\mathcal{X}$ such that $R_i(\vx)\geq r_i^0$ for every user with $w_i>0$, then $D^\star=0$ and hence $D(\vx_\lambda)\leq \frac{2M_S}{\lambda}$. If, in addition, for each user with $w_i>0$ there exists $c_i>0$ such that $d_i(t)\geq c_i t,
\qquad t\geq0,$ then every positively weighted user satisfies
\begin{equation}\label{eq:individual_qos_deficit_bound}
\bigl(r_i^0-R_i(\vx_\lambda)\bigr)_+\leq \frac{2M_S}{\lambda w_i c_i},\qquad w_i>0.
\end{equation}
Thus,
\begin{equation}
D(\vx_\lambda)-D^\star = O(1/\lambda).
\end{equation}
When the QoS targets of all positively weighted users are
jointly feasible, $D^\star=0$ and the aggregate soft-QoS
deficit measure $D(\vx_\lambda)$ is itself $O(1/\lambda)$.
Under the stated linear growth condition on $d_i$, each
positively weighted user's raw QoS shortfall is also
$O(1/\lambda)$.
\end{corollary}

\begin{proof}
Fix $\lambda>0$ and let $\vx_\lambda\in
\arg\max_{\vx\in\mathcal{X}}F_\lambda(\vx)$.
Let $\vx^D\in\arg\min_{\vx\in\mathcal{X}}D(\vx)$.
By global optimality of $\vx_\lambda$, 
\begin{equation}
S(\vx_\lambda)-\lambda D(\vx_\lambda)
\geq
S(\vx^D)-\lambda D^\star.
\end{equation}
Therefore,
\begin{equation}
0
\leq
D(\vx_\lambda)-D^\star
\leq
\frac{S(\vx_\lambda)-S(\vx^D)}{\lambda}
\leq
\frac{2M_S}{\lambda},
\end{equation}
which proves \eqref{eq:soft_qos_rate_cor}.

If there exists an allocation satisfying
$R_i(\vx)\geq r_i^0$ for every $i$ with $w_i>0$, then
$D^\star=0$. Hence,
\begin{equation}
D(\vx_\lambda)\leq \frac{2M_S}{\lambda}.
\end{equation}

Now fix any user $i$ with $w_i>0$. Under the growth condition
$d_i(t)\geq c_i t$,
\begin{equation}
D(\vx_\lambda)
=
\sum_{j=1}^N
w_j d_j\!\left(
(r_j^0-R_j(\vx_\lambda))_+
\right)
\geq
w_i c_i
(r_i^0-R_i(\vx_\lambda))_+.
\end{equation}
Combining this inequality with
$D(\vx_\lambda)\leq 2M_S/\lambda$ gives
\begin{equation}
(r_i^0-R_i(\vx_\lambda))_+
\leq
\frac{2M_S}{\lambda w_i c_i},
\end{equation}
which proves \eqref{eq:individual_qos_deficit_bound}.
\end{proof}

\begin{remark}[Global versus stationary penalized solutions]\label{rem:soft_qos_global_stationary}
Theorem~\ref{thm:soft_qos} and Corollary~\ref{cor:soft_qos_rate} concern global maximizers of the penalized problem, and do not follow from first-order stationarity. Consequently, for a nonconcave penalized objective, an SCA iterate inherits the stationary-point guarantee shown in Section~\ref{sec:sca}, but not automatically the $O(1/\lambda)$ global excess-deficit bound (or the $O(1/\lambda)$ deficit bound when $D^\star=0$).
\end{remark}

The utility constructions above lead to objectives that may be nonconcave, nonsmooth, or both. The next section therefore develops a maximization-consistent SCA framework for solving the resulting allocation problems.

\section{Successive Convex Approximation}\label{sec:sca}
The considered allocation problems may be nonconvex because of interference coupling, nonconcave reference-dependent utilities, nonlinear probability-weighted objectives, and compositions of nonlinear functions. We use SCA to solve a sequence of tractable surrogate problems. Because the original allocation problems are formulated as maximization problems, each surrogate must be consistent with the maximization form of the objective.

The proposed SCA framework is structure exploiting rather than a generic black-box method for nonsmooth nonconvex optimization. Generic complexity results for arbitrary Lipschitz nonsmooth nonconvex problems therefore do not directly characterize the structured resource-allocation problems considered here \cite{JordanNonsmoothNonconvex2023}. We instead exploit the separability of user utilities, the structure of communication reward functions, and explicit smooth/nonsmooth decompositions of reference-dependent utilities.

The role of SCA here is algorithmic: it replaces a difficult nonconvex allocation problem with a sequence of local convex or concave surrogate problems while preserving feasibility with respect to the current convex constraint set. In contrast to generic black-box nonlinear programming, the surrogate construction can exploit the structure of the communication rate expression, the separability of the utilities, and the simplex geometry of power constraints.

We first state the generic constrained SCA form used throughout the deterministic and stochastic resource-allocation problems. This formulation makes explicit how feasibility is preserved when nonconvex constraints are included.

\subsection{Constrained SCA Model and Surrogate Requirements}
Consider the nonconvex maximization problem
\begin{equation}\label{eq:generic_constrained_sca_problem}
\begin{aligned}
    \max_{\vx}\quad & F(\vx)\\
    \text{s.t.}\quad & g_m(\vx)\leq 0,\quad m=1,\ldots,M,\\
    & \vx\in\calS,
\end{aligned}
\end{equation}
where $\calS\subseteq\R^n$ is nonempty, closed, and convex. In the power allocation problems considered in this paper, the feasible set is compact because of the total power constraint; therefore, the coercivity condition sometimes imposed in generic SCA statements is not required. At iteration $l$, the original problem is approximated by the surrogate problem
\begin{equation}\label{eq:generic_constrained_sca_subproblem}
\begin{aligned}
    \max_{\vx}\quad & \widetilde F(\vx;\vx^{(l)})\\
    \text{s.t.}\quad & \widetilde g_m(\vx;\vx^{(l)})\leq 0,\quad m=1,\ldots,M,\\
    & \vx\in\calS.
\end{aligned}
\end{equation}

For the maximization problem, the objective surrogate $\widetilde F(\cdot;\vx^{(l)})$ should be concave in $\vx$. The convergence result in Section~V-B additionally assumes uniform strong concavity, which also guarantees uniqueness of the surrogate best response. The surrogate should satisfy
\begin{subequations}\label{eq:objective_surrogate_conditions}
\begin{align}
\widetilde F(\vx^{(l)};\vx^{(l)})
    &= F(\vx^{(l)}),\\
    \nabla_{\vx}\widetilde F(\vx^{(l)};\vx^{(l)})
    &= \nabla F(\vx^{(l)}),\\
    \widetilde F(\vx;\vy)
    \ \text{and}\
    \nabla_{\vx}\widetilde F(\vx;\vy)
    &\ \text{are jointly continuous in }(\vx,\vy).
\end{align}
\end{subequations}
If a constraint $g_m$ is nonconvex, a feasibility preserving convex majorant is used:
\begin{subequations}\label{eq:constraint_surrogate_conditions}
\begin{align}
    \widetilde g_m(\cdot;\vx^{(l)})&\;\text{is convex},\\
    \widetilde g_m(\vx^{(l)};\vx^{(l)})&=g_m(\vx^{(l)}),\\
    g_m(\vx)&\leq \widetilde g_m(\vx;\vx^{(l)}),\quad \forall \vx\in\calS,\\
    \nabla_{\vx}\widetilde g_m(\vx^{(l)};\vx^{(l)})&=\nabla g_m(\vx^{(l)}).
\end{align}
\end{subequations}
The upper-bound condition in \eqref{eq:constraint_surrogate_conditions} ensures that every point feasible for the surrogate subproblem is also feasible for the corresponding original constraint over the approximation domain. Moreover, by the touching condition, a feasible current iterate $\vx^{(l)}$ is feasible for the surrogate subproblem. Hence, a full-step update $\vx^{(l+1)}=\widehat{\vx}^{(l)}$
preserves feasibility. For relaxed updates, feasibility must be ensured separately unless additional structure guarantees that the update remains feasible.

\subsection{Smooth Maximization SCA}
Consider
\begin{equation}\label{eq:smooth_nonconvex}
    \max_{\vx\in\calX} F(\vx),
\end{equation}
where $\calX$ is nonempty, compact, and convex, and $F$ is continuously differentiable with a Lipschitz-continuous gradient. A simple maximization-consistent surrogate is
\begin{equation}\label{eq:proximal_surrogate}
\widetilde F(\vx;\vy)=F(\vy)+\nabla F(\vy)^T(\vx-\vy)-\frac{\rho}{2}\|\vx-\vy\|_2^2, \ \rho>0.
\end{equation}
The negative quadratic term makes the surrogate strongly concave in $\vx$. Theorem~\ref{thm:sca_convergence} below is stated for a fixed surrogate map $\widetilde F(\vx;\vy)$; hence, when the proximal form in \eqref{eq:proximal_surrogate} is used, $\rho$ is held fixed across outer iterations. Allowing an iteration-dependent $\rho_l$ requires an explicitly iteration-varying surrogate analysis and is not covered by the fixed-best-response-map proof used here.

Here and throughout, an accumulation point denotes the limit of a convergent subsequence.

\begin{algorithm}[t]
\caption{Maximization-consistent SCA}
\label{alg:SCA}
\begin{algorithmic}[1]
    \STATE Set $l=0$ and choose a feasible point $\vx^{(0)}\in\calX$.
    \REPEAT
    \STATE Compute $\widehat{\vx}^{(l)}\in\argmax_{\vx\in\calX}\widetilde F(\vx;\vx^{(l)})$.
    \STATE Set $\vx^{(l+1)}=\vx^{(l)}+\gamma_l(\widehat{\vx}^{(l)}-\vx^{(l)})$.
    \STATE $l\leftarrow l+1$.
    \UNTIL convergence or a stopping criterion is met
    \STATE \textbf{return} $\vx^{(l)}$.
\end{algorithmic}
\end{algorithm}

\begin{theorem}[Stationarity of SCA accumulation points, \cite{SCA_Scutari_Theory_Part1}]\label{thm:sca_convergence}
    Assume $\calX$ is nonempty, compact, and convex and $F$ is continuously differentiable with a Lipschitz-continuous gradient on an open set containing $\calX$. Assume further that a single, iteration-independent surrogate map $\widetilde F(\vx;\vy)$ is used throughout the run, that it is jointly continuous in $(\vx,\vy)$, continuously differentiable in its first argument, and that $\nabla_{\vx}\widetilde F(\vx;\vy)$ is jointly continuous. Suppose $\widetilde F(\cdot;\vy)$ is uniformly strongly concave with modulus $\sigma>0$ independent of $\vy\in\calX$, and is value- and gradient-consistent, i.e., $\widetilde F(\vy;\vy)=F(\vy)$ and $\nabla_{\vx}\widetilde F(\vy;\vy)=\nabla F(\vy)$. If every surrogate problem is solved exactly and the step sizes satisfy $\gamma_l\in(0,1]$, $\gamma_l\to0$, $\sum_l\gamma_l=\infty$, and $\sum_l\gamma_l^2<\infty$, then every accumulation point of Algorithm~\ref{alg:SCA} is a first-order stationary point of \eqref{eq:smooth_nonconvex}.
\end{theorem}
\begin{proof}
    Define the single-valued best-response map
    \begin{equation*}
        \mathcal B(\vy)\triangleq
        \argmax_{\vx\in\calX}\widetilde F(\vx;\vy),
        \qquad
        \vd(\vy)\triangleq\mathcal B(\vy)-\vy,
    \end{equation*}
    where uniqueness follows from uniform strong concavity. By Berge's maximum theorem and uniqueness, $\mathcal B$ and $\vd$ are continuous on the compact set $\calX$.

    Let $\widehat\vx^{(l)}=\mathcal B(\vx^{(l)})$ and $\vd^{(l)}=\widehat\vx^{(l)}-\vx^{(l)}$. Strong concavity with modulus $\sigma$ gives
    \begin{align*}
        \widetilde F(\widehat\vx^{(l)};\vx^{(l)})
        &\leq \widetilde F(\vx^{(l)};\vx^{(l)})
        +\nabla_{\vx}\widetilde F(\vx^{(l)};\vx^{(l)})^T\vd^{(l)} \\
        & \quad -\frac{\sigma}{2}\|\vd^{(l)}\|_2^2.
    \end{align*}
    Since $\widehat\vx^{(l)}$ maximizes the surrogate,
    $\widetilde F(\widehat\vx^{(l)};\vx^{(l)})\geq
    \widetilde F(\vx^{(l)};\vx^{(l)})$. Using gradient consistency therefore yields
    \begin{equation}
        \nabla F(\vx^{(l)})^T\vd^{(l)}
        \geq \frac{\sigma}{2}\|\vd^{(l)}\|_2^2.
        \label{eq:sca_gradient_related}
    \end{equation}
    Let $L_F$ be a Lipschitz constant of $\nabla F$ on $\calX$. The ascent lemma and the update
    $\vx^{(l+1)}=\vx^{(l)}+\gamma_l\vd^{(l)}$ imply
    \begin{align}
    F(\vx^{(l+1)})
        &\geq F(\vx^{(l)})
        +\gamma_l\nabla F(\vx^{(l)})^T\vd^{(l)}
        -\frac{L_F}{2}\gamma_l^2\|\vd^{(l)}\|_2^2\notag\\
        &\geq F(\vx^{(l)})
        +\frac{\gamma_l}{2}(\sigma-L_F\gamma_l) \|\vd^{(l)}\|_2^2.
        \label{eq:sca_ascent_bound}
    \end{align}
    Because $\gamma_l\to0$, there exists $l_0$ such that
    $L_F\gamma_l\leq\sigma/2$ for all $l\geq l_0$. Hence
    \begin{equation}
        F(\vx^{(l+1)})-F(\vx^{(l)})
        \geq \frac{\sigma}{4}\gamma_l\|\vd^{(l)}\|_2^2,
        \qquad l\geq l_0.
        \label{eq:sca_sufficient_ascent}
    \end{equation}
    Since $F$ is bounded above on compact $\calX$, summing
    \eqref{eq:sca_sufficient_ascent} gives
    \begin{equation}
        \sum_{l=l_0}^{\infty}\gamma_l\|\vd^{(l)}\|_2^2<\infty.
        \label{eq:sca_weighted_residual}
    \end{equation}
    Moreover, compactness gives $\|\vd^{(l)}\|_2\leq\operatorname{diam}(\calX)$, and therefore
    \begin{equation*}
        \|\vx^{(l+1)}-\vx^{(l)}\|_2 =\gamma_l\|\vd^{(l)}\|_2\longrightarrow0.
    \end{equation*}

We next show that every accumulation point is a fixed point of $\mathcal B$. Suppose, to the contrary, that a subsequence
$\vx^{(l_j)}\to\bar\vx$ satisfies
$\|\vd(\bar\vx)\|_2>0$. By continuity of $\vd$, there exist
$r>0$ and $\varepsilon>0$ such that
\begin{equation*}
\|\vd(\vx)\|_2\geq\varepsilon,
\qquad \vx\in\calX\cap B(\bar\vx,r).
\end{equation*}
For all sufficiently large $j$, the iterate $\vx^{(l_j)}$ lies in
$B(\bar\vx,r/2)$. If the sequence remains in $B(\bar\vx,r)$ forever, then \eqref{eq:sca_weighted_residual} and
$\sum_l\gamma_l=\infty$ contradict the lower bound
$\|\vd^{(l)}\|_2\geq\varepsilon$. Otherwise, let $m_j>l_j$ be the first index leaving $B(\bar\vx,r)$. With
$D_{\calX}\triangleq\operatorname{diam}(\calX)$,
\begin{align*}
\frac{r}{2}
&\leq\|\vx^{(m_j)}-\vx^{(l_j)}\|_2
\leq\sum_{l=l_j}^{m_j-1}\gamma_l\|\vd^{(l)}\|_2
\leq D_{\calX}\sum_{l=l_j}^{m_j-1}\gamma_l.
\end{align*}
Thus each such visit contributes at least
$r/(2D_{\calX})$ to the partial sum of $\gamma_l$. Using
\eqref{eq:sca_sufficient_ascent} and
$\|\vd^{(l)}\|_2\geq\varepsilon$ before exit gives
\begin{equation*}
F(\vx^{(m_j)})-F(\vx^{(l_j)})
\geq \frac{\sigma\varepsilon^2r}{8D_{\calX}}>0.
\end{equation*}
If exits occur infinitely often, the accumulating subsequence contains infinitely many returns to
$B(\bar\vx,r/2)$; hence one can select a subsequence of entry--exit pairs $[l_{j_k},m_{j_k}]$ that are pairwise disjoint. Since \eqref{eq:sca_sufficient_ascent} makes $F(\vx^{(l)})$ nondecreasing for all sufficiently large $l$, each such disjoint cycle increases $F$ by at least
$\sigma\varepsilon^2r/(8D_{\calX})$. Infinitely many such cycles would therefore force $F(\vx^{(l)})\to+\infty$, contradicting boundedness. Hence
$\mathcal B(\bar\vx)=\bar\vx$.

Finally, first-order optimality of the concave surrogate at its fixed point gives
    \begin{equation*}
        \nabla_{\vx}\widetilde F(\bar\vx;\bar\vx)^T (\vx-\bar\vx)\leq0, \qquad \forall\vx\in\calX.
    \end{equation*}
    Gradient consistency yields
    \begin{equation*}
        \nabla F(\bar\vx)^T(\vx-\bar\vx)\leq0, \qquad \forall\vx\in\calX,
    \end{equation*}
    which is equivalent to $0\in-\nabla F(\bar\vx)+N_{\calX}(\bar\vx)$.
\end{proof}

\begin{corollary}[Finite-time smooth-SCA residual bound]\label{cor:sca_finite_time}
Under the smooth problem and surrogate assumptions of Theorem~\ref{thm:sca_convergence}, replace its diminishing-step condition by a constant step size $\gamma\in(0,1]$ satisfying $\sigma-L_F\gamma>0$, where $L_F$ is a Lipschitz constant of $\nabla F$ on $\calX$. Let $F_{\max}=\max_{\vx\in\calX}F(\vx)$ and $\vd^{(l)}=\mathcal B(\vx^{(l)})-\vx^{(l)}$. Then, for every integer $T\geq1$,
\begin{equation}\label{eq:sca_finite_time_bound}
\min_{0\leq l<T}\|\vd^{(l)}\|_2^2\leq\frac{2\bigl(F_{\max}-F(\vx^{(0)})\bigr)}{\gamma(\sigma-L_F\gamma)T}.
\end{equation}
Consequently, an iterate with $\|\vd^{(l)}\|_2\leq\epsilon$ is obtained within $O(\epsilon^{-2})$ outer iterations. Moreover, $\|\vd^{(l)}\|_2\to0$, so every accumulation point is first-order stationary for \eqref{eq:smooth_nonconvex}. 
Under the fixed-surrogate and strong-concavity assumptions above, the best-response residual is zero if and only if the current point is first-order stationary.
\end{corollary}
\begin{proof}
With $\gamma_l\equiv\gamma$, inequality~\eqref{eq:sca_ascent_bound} gives
\begin{equation*}
F(\vx^{(l+1)})-F(\vx^{(l)})\geq\frac{\gamma}{2}(\sigma-L_F\gamma)\|\vd^{(l)}\|_2^2.
\end{equation*}
Summing from $l=0$ to $T-1$ and using $F(\vx^{(T)})\leq F_{\max}$ yields \eqref{eq:sca_finite_time_bound}. Summing the same ascent inequality over all $l$ shows $\sum_l\|\vd^{(l)}\|_2^2<\infty$, hence $\|\vd^{(l)}\|_2\to0$. If $\vx^{(l_j)}\to\bar\vx$ is any convergent subsequence, continuity of $\vd(\vx)=\mathcal B(\vx)-\vx$ gives $\vd(\bar\vx)=\lim_{j\to\infty}\vd^{(l_j)}=0$; hence every accumulation point is a fixed point of $\mathcal B$. If $\vd(\vx)=0$, first-order optimality of the concave surrogate and gradient consistency give stationarity of $F$. Conversely, if $\vx$ is stationary for $F$, gradient consistency makes $\vx$ satisfy the first-order optimality condition of the strongly concave surrogate $\widetilde F(\cdot;\vx)$, so $\mathcal B(\vx)=\vx$.
\end{proof}

For the constrained problem \eqref{eq:generic_constrained_sca_problem}, an analogous conclusion requires uniformly regular objective and constraint surrogates, nonempty inner-approximation feasible sets, a feasible initialization, and an appropriate constraint qualification at every accumulation point. Under these additional hypotheses, feasible iterates are preserved and each accumulation point satisfies the KKT conditions of the original constrained problem. Consistency alone does not guarantee ascent at every iteration. In the smooth case, however, Corollary~\ref{cor:sca_finite_time} shows that a sufficiently small constant step ensures a positive ascent bound; line searches or genuine lower-bound/minorization surrogates are alternative mechanisms for enforcing monotonic improvement.

\subsection{Dual Projected-Subgradient Implementation}\label{subsec:dual_projected_subgradient}
When the SCA subproblem is separable across users except for an aggregate resource constraint, Lagrangian relaxation decouples the primal variables and yields a one-dimensional dual problem that can be solved by a projected-subgradient update. This approach is applicable even when the primal surrogate is smooth, because the resulting dual function may be nonsmooth.

For separable power-allocation subproblems, the SCA surrogate often has the form
\begin{equation}\label{eq:separable_sca_subproblem}
\begin{aligned}
    \max_{\vP}\quad & \sum_{i=1}^{N}\widetilde \phi_i(P_i;\vP^{(l)})\\
    \text{s.t.}\quad & P_i\geq0,\quad i=1,\ldots,N,\\
    & \sum_{i=1}^{N}P_i\leq P_{\rm tot},
\end{aligned}
\end{equation}
where each $\widetilde \phi_i$ is strongly concave. This structure was exploited in the conference algorithm through Lagrangian relaxation and a projected-subgradient update \cite{vaidanis2025SPAWC}. 

Recent Lagrangian-based frameworks for nonsmooth nonconvex constrained optimization also support the use of subgradient-type primal updates together with dual updates under conservative-Jacobian or path-differentiability assumptions \cite{XiaoLagrangianNonsmooth2024}. In the present work, however, the Lagrangian relaxation is applied to the convexified SCA subproblem; hence, the convergence guarantee remains governed by the outer SCA assumptions and by the accuracy of the inner solver.

Define
\begin{equation}\label{eq:dual_lagrangian_max}
    \mathcal L(\vP,\lambda)=\sum_{i=1}^{N}\widetilde \phi_i(P_i;\vP^{(l)})-\lambda\left(\sum_{i=1}^{N}P_i-P_{\rm tot}\right),\quad \lambda\geq0.
\end{equation}
For a fixed $\lambda$, the maximizer separates across users:
\begin{equation}\label{eq:separable_primal_recovery}
    P_i^{\star}(\lambda)\in\argmax_{P_i\in[0,P_{\rm tot}]}\;\widetilde \phi_i(P_i;\vP^{(l)})-\lambda P_i.
\end{equation}
The bounds $P_i\leq P_{\rm tot}$ are redundant in the primal simplex but are retained in the scalar dual maximizations to make the dual subgradients uniformly bounded. The dual function is $q(\lambda)=\max_{\vP\in[0,P_{\rm tot}]^N}\mathcal L(\vP,\lambda)$, and the dual problem is $\min_{\lambda\geq0}q(\lambda)$. A projected-subgradient step is
\begin{equation}\label{eq:dual_update}
\lambda^{(t+1)}=\left[\lambda^{(t)}+\zeta_t\left(\sum_{i=1}^{N}P_i^{\star}(\lambda^{(t)})-P_{\rm tot}\right)\right]_+,
\end{equation}
where $\zeta_t>0$ is square-summable but not summable. If \eqref{eq:separable_primal_recovery} has no closed-form solution, it can be solved by projected-gradient ascent on the same compact scalar domain $P_i\in[0,P_{\rm tot}]$.

\begin{proposition}[Dual convergence for the simplex SCA subproblem]\label{prop:dual_convergence}
Assume $P_{\rm tot}>0$ and that each $\widetilde\phi_i$ is proper, closed, continuously differentiable, and uniformly strongly concave on the relevant compact interval $[0,P_{\rm tot}]$. Then \eqref{eq:separable_sca_subproblem} has a unique primal optimum and strong duality holds. For step sizes satisfying $\sum_t\zeta_t=\infty$ and $\sum_t\zeta_t^2<\infty$, the projected iterates in \eqref{eq:dual_update} converge to a dual-optimal point; in particular, their distance to the dual optimal set converges to zero. The recovered sequence $\vP^\star(\lambda^{(t)})$ converges to the unique primal optimum.
\end{proposition}

\begin{proof}
Add the redundant box constraints
$0\leq P_i\leq P_{\rm tot}$. The feasible set is compact and has a Slater point, for example
\begin{equation*}
P_i=\frac{P_{\rm tot}}{2N},
\qquad i=1,\ldots,N,
\end{equation*}
for which $\sum_iP_i=P_{\rm tot}/2<P_{\rm tot}$. After sign reversal, \eqref{eq:separable_sca_subproblem} is a convex program; Slater's condition implies zero duality gap and existence of a dual optimum. If each $\widetilde\phi_i$ is strongly concave with modulus at least $m>0$, then
\begin{align*}
&\sum_{i=1}^{N}\widetilde\phi_i
\bigl(\theta P_i+(1-\theta)Q_i\bigr)\\
&\quad\geq \theta\sum_{i=1}^{N}\widetilde\phi_i(P_i)\\
&\qquad +(1-\theta)\sum_{i=1}^{N}\widetilde\phi_i(Q_i)\\
&\qquad +\frac{m}{2}\theta(1-\theta)
\|\vP-\vQ\|_2^2,
\end{align*}
so the primal optimizer $\vP^\star$ is unique.

For fixed $\lambda$, the Lagrangian is separable and the unique maximizer is
$\vP^\star(\lambda)=(P_1^\star(\lambda),\ldots,P_N^\star(\lambda))$. Danskin's theorem gives
\begin{equation}
q'(\lambda)=P_{\rm tot}-\sum_{i=1}^{N}P_i^\star(\lambda)
\triangleq g(\lambda).
\label{eq:dual_gradient_danskin}
\end{equation}
Thus \eqref{eq:dual_update} is exactly
\begin{equation*}
\lambda^{(t+1)}=\Pi_{\R_+}
\left(\lambda^{(t)}-\zeta_t g(\lambda^{(t)})\right).
\end{equation*}
Under the present strong-concavity assumptions,
$q$ is differentiable, so this is a projected-gradient step for the scalar convex dual problem (and is, in particular, a valid projected-subgradient step). Because $0\leq P_i^\star(\lambda)\leq P_{\rm tot}$,
\begin{equation*}
|g(\lambda)|\leq (N+1)P_{\rm tot}\triangleq G.
\end{equation*}
Let $\lambda^\star$ be any dual optimum. Nonexpansiveness of projection and the subgradient inequality for convex $q$ yield
\begin{equation} \label{eq:dual_quasi_fejer}
    \begin{split}
        |\lambda^{(t+1)}-\lambda^\star|^2 &\leq |\lambda^{(t)}-\lambda^\star|^2 -2\zeta_t g(\lambda^{(t)})(\lambda^{(t)}-\lambda^\star) \\
        & \quad+\zeta_t^2G^2 \\
        & \leq |\lambda^{(t)}-\lambda^\star|^2 -2\zeta_t\bigl(q(\lambda^{(t)})-q^\star\bigr) +\zeta_t^2G^2.
    \end{split}
\end{equation}
Since $\sum_t\zeta_t^2<\infty$, \eqref{eq:dual_quasi_fejer} is a quasi-Fej\'er inequality with respect to the nonempty closed dual optimal set $\Lambda^\star$. Summing it gives
\begin{equation*}
\sum_{t=0}^{\infty}\zeta_t
\bigl(q(\lambda^{(t)})-q^\star\bigr)<\infty.
\end{equation*}
Together with $\sum_t\zeta_t=\infty$, this implies
$\liminf_t q(\lambda^{(t)})=q^\star$. The quasi-Fej\'er inequality also implies that $\{\lambda^{(t)}\}$ is bounded and that, for every $\lambda^\star\in\Lambda^\star$, the sequence
$|\lambda^{(t)}-\lambda^\star|$ has a limit. Hence there exists a subsequence $\lambda^{(t_k)}\to\bar\lambda$ along which $q(\lambda^{(t_k)})\to q^\star$. By continuity of $q$, $\bar\lambda\in\Lambda^\star$. Applying the preceding distance limit property with $\lambda^\star=\bar\lambda$ and using $\lambda^{(t_k)}\to\bar\lambda$ shows that the corresponding limit is zero. Therefore
\begin{equation*}
\lambda^{(t)}\longrightarrow\bar\lambda\in\Lambda^\star,
\end{equation*} and, in particular,
\begin{equation*}
\operatorname{dist}(\lambda^{(t)},\Lambda^\star)\longrightarrow0.
\end{equation*}

Finally, uniform strong concavity and joint continuity imply continuity of the best-response map $\lambda\mapsto\vP^\star(\lambda)$. For every $\bar\lambda\in\Lambda^\star$, strong duality and primal uniqueness give
\begin{equation*}
\vP^\star(\bar\lambda)=\vP^\star.
\end{equation*}
Since $\lambda^{(t)}\to\bar\lambda$ and
$\vP^\star(\cdot)$ is continuous, it follows directly that
\begin{equation*}
\vP^\star(\lambda^{(t)})\longrightarrow\vP^\star.
\end{equation*}
\end{proof}

\subsection{Nonsmooth Reference Points}
Reference-dependent utilities may have a kink at the reference point. For a utility with a single downward kink that can be isolated by a hinge term, a useful decomposition is
\begin{equation}\label{eq:nonsmooth_decomposition}
    u_i(r)=s_i(r)-\kappa_i(r-r_i^0)_+,
\end{equation}
where $s_i$ is smooth and $\kappa_i\geq0$. For a piecewise-smooth utility with finite one-sided derivatives, the natural choice is $\kappa_i=u_i'(r_i^{0-})-u_i'(r_i^{0+})\geq0$, provided that the resulting $s_i$ satisfies the stated smoothness assumptions. The term $-(r-r_i^0)_+$ is concave, so it can be retained exactly in a maximization surrogate, whereas the smooth, possibly nonconcave part can be linearized and regularized.

For a scalar orchestrator utility $U_{\rm orch}$ with reference point $z_0$, left derivative $D^-U_{\rm orch}(z_0)$, and right derivative $D^+U_{\rm orch}(z_0)$, the decomposition used in the stochastic formulation can be written explicitly as
\begin{subequations}\label{eq:orch_decomposition}
\begin{equation}
    \begin{split}
        \widehat U_{\rm orch}(z) &=U_{\rm orch}(z) \\
        & \quad +\left(D^-U_{\rm orch}(z_0)-D^+U_{\rm orch}(z_0)\right)(z-z_0)_+,
    \end{split}
\end{equation}
\begin{equation}
    \begin{split}
        \widetilde U_{\rm orch}(z) &=\left(D^+U_{\rm orch}(z_0)-D^-U_{\rm orch}(z_0)\right)(z-z_0)_+.
    \end{split}
\end{equation}
\end{subequations}
Then $U_{\rm orch}=\widehat U_{\rm orch}+\widetilde U_{\rm orch}$. If $D^+U_{\rm orch}(z_0)\leq D^-U_{\rm orch}(z_0)$, which corresponds to a downward kink at the reference point, $\widehat U_{\rm orch}$ has equal one-sided slopes at $z_0$, while $\widetilde U_{\rm orch}$ is concave. Thus, the nonsmooth concave component can be kept exactly in the surrogate, whereas the remaining smooth, possibly nonconcave component is handled by the proximal SCA step. Slope matching at $z_0$ is only a first-order statement. Proposition~\ref{prop:nonsmooth_sca} applies only when the corrected component $\widehat U_{\rm orch}$ (after composition with the scenario reward) is continuously differentiable with a Lipschitz-continuous gradient on the relevant compact domain. In particular, value functions with an unbounded derivative at the reference point are outside this smooth-SCA assumption unless they are regularized or treated by a different nonsmooth analysis.

\begin{proposition}[Composite nonsmooth SCA stationarity]\label{prop:nonsmooth_sca}
Let $F(\vx)=F_s(\vx)+F_c(\vx)$ on a nonempty, compact, convex set $\calX$, where $F_s$ is continuously differentiable with Lipschitz-continuous gradient and $F_c$ is finite, concave, and locally Lipschitz on a neighborhood of $\calX$. Suppose the smooth surrogate for $F_s$ satisfies the uniform assumptions of Theorem~\ref{thm:sca_convergence}, while $F_c$ is kept exactly, and suppose the subproblems are solved exactly with the same step-size conditions. Then every accumulation point $\bar\vx$ is Clarke-stationary in the constrained maximization sense,
\begin{equation*}
0\in-\nabla F_s(\bar\vx)-\partial_C F_c(\bar\vx)+N_{\calX}(\bar\vx),
\end{equation*}
where $\partial_C$ is the Clarke generalized gradient and $N_{\calX}$ is the convex normal cone.
\end{proposition}
\begin{proof}
Define the composite surrogate and best-response map
\begin{align*}
H(\vx;\vy)&\triangleq\widetilde F_s(\vx;\vy)+F_c(\vx),\\
\mathcal B_c(\vy)&\triangleq\argmax_{\vx\in\calX}H(\vx;\vy),\\
\vd_c(\vy)&\triangleq\mathcal B_c(\vy)-\vy.
\end{align*}
Because $\widetilde F_s(\cdot;\vy)$ is uniformly strongly concave and $F_c$ is concave, $H(\cdot;\vy)$ is uniformly strongly concave; hence $\mathcal B_c$ is single-valued. Moreover, $F_c$ is continuous on $\calX$ and $\widetilde F_s$ is jointly continuous, so the maximum theorem together with uniqueness of the maximizer implies that $\mathcal B_c$ is continuous. Let
$\widehat\vx^{(l)}=\mathcal B_c(\vx^{(l)})$ and
$\vd_c^{(l)}=\widehat\vx^{(l)}-\vx^{(l)}$. Strong concavity and surrogate optimality give
\begin{equation}
\nabla F_s(\vx^{(l)})^T\vd_c^{(l)}
 +F_c(\widehat\vx^{(l)})-F_c(\vx^{(l)})
\geq\frac{\sigma}{2}\|\vd_c^{(l)}\|_2^2.
\label{eq:composite_gradient_related}
\end{equation}
Indeed, this follows by applying the strong-concavity inequality to
$H(\cdot;\vx^{(l)})$ at $\vx^{(l)}$ and using
$H(\widehat\vx^{(l)};\vx^{(l)})\geq
H(\vx^{(l)};\vx^{(l)})$.

For $\vx^{(l+1)}=(1-\gamma_l)\vx^{(l)}+
\gamma_l\widehat\vx^{(l)}$, concavity of $F_c$ gives
\begin{equation*}
F_c(\vx^{(l+1)})
\geq(1-\gamma_l)F_c(\vx^{(l)})
 +\gamma_lF_c(\widehat\vx^{(l)}).
\end{equation*}
Combining this with the ascent lemma for $F_s$ and
\eqref{eq:composite_gradient_related} yields
\begin{equation}
\begin{aligned}
F(\vx^{(l+1)})-F(\vx^{(l)})
&\geq \gamma_l\Bigl[
\nabla F_s(\vx^{(l)})^T\vd_c^{(l)} \\
&\qquad
+F_c(\widehat\vx^{(l)})
-F_c(\vx^{(l)})
\Bigr] \\
&\quad
-\frac{L_s}{2}\gamma_l^2\|\vd_c^{(l)}\|_2^2 \\
&\geq
\frac{\gamma_l}{2}(\sigma-L_s\gamma_l)
\|\vd_c^{(l)}\|_2^2.
\end{aligned}
\end{equation}
The same summability and neighborhood-exit argument used in the proof of Theorem~\ref{thm:sca_convergence} therefore gives
\begin{equation*}
\mathcal B_c(\bar\vx)=\bar\vx
\end{equation*}
for every accumulation point $\bar\vx$.

At such a fixed point, the first-order condition for maximizing the concave function $H(\cdot;\bar\vx)$ over $\calX$, written in sign-reversed convex form, is
\begin{equation*}
0\in-
\nabla_{\vx}\widetilde F_s(\bar\vx;\bar\vx)
-\partial_C F_c(\bar\vx)+N_{\calX}(\bar\vx).
\end{equation*}
Using gradient consistency,
$\nabla_{\vx}\widetilde F_s(\bar\vx;\bar\vx)=
\nabla F_s(\bar\vx)$, gives
\begin{equation*}
0\in-\nabla F_s(\bar\vx)-\partial_C F_c(\bar\vx)
+N_{\calX}(\bar\vx).
\end{equation*}
Since $F_s$ is continuously differentiable, the Clarke sum rule gives $\partial_C(F_s+F_c)(\bar\vx)=
\nabla F_s(\bar\vx)+\partial_C F_c(\bar\vx)$. Thus, the preceding inclusion is precisely the constrained
Clarke-stationarity condition to maximize $F=F_s+F_c$.
\end{proof}

\subsection{Interference-Limited Rate Surrogates}
For the interference-limited rate in \eqref{eq:interference_rate}, define
\begin{align}      a_i(\vP)&=\log_2\!\left(N_0+\sum_{j=1}^{N}g_{ij}P_j\right),\\
b_i(\vP)&=\log_2\!\left(N_0+\sum_{j\neq i}g_{ij}P_j\right).
\end{align}
Then $R_i(\vP) = a_i(\vP) - b_i(\vP)$, where both $a_i$ and $b_i$ are concave.

\begin{proposition}[Concave lower bound for interference rates]\label{prop:dc_rate}
At iteration $l$, the function
\begin{equation}\label{eq:rate_lower_bound}
    \widehat R_i(\vP;\vP^{(l)})=a_i(\vP)-b_i(\vP^{(l)})-\nabla b_i(\vP^{(l)})^T(\vP-\vP^{(l)})
\end{equation}
is a concave lower bound of $R_i(\vP)$ over the feasible power domain and is tight at $\vP=\vP^{(l)}$. Moreover, it is gradient-consistent at the expansion point.
\end{proposition}
\begin{proof}
The first-order characterization of concavity gives, for every $\vP$,
\begin{equation}
b_i(\vP)\leq b_i(\vP^{(l)})
+\nabla b_i(\vP^{(l)})^T(\vP-\vP^{(l)}).
\label{eq:bi_tangent_upper_bound}
\end{equation}
Hence
\begin{align*}
R_i(\vP)-\widehat R_i(\vP;\vP^{(l)})
&=b_i(\vP^{(l)})
+\nabla b_i(\vP^{(l)})^T(\vP-\vP^{(l)}) \\
& \quad -b_i(\vP) \geq0,
\end{align*}
which proves the lower-bound property. Since the last two terms in
\eqref{eq:rate_lower_bound} are affine in $\vP$ and $a_i$ is concave,
$\widehat R_i(\cdot;\vP^{(l)})$ is concave. At the expansion point,
\begin{align*}
\widehat R_i(\vP^{(l)};\vP^{(l)})
&=a_i(\vP^{(l)})-b_i(\vP^{(l)})
=R_i(\vP^{(l)}),\\
\nabla_{\vP}\widehat R_i(\vP^{(l)};\vP^{(l)})
&=\nabla a_i(\vP^{(l)})-\nabla b_i(\vP^{(l)})
=\nabla R_i(\vP^{(l)}).
\end{align*}
Thus the lower bound is both value- and gradient-consistent.
\end{proof}
Note that if strong concavity of the surrogate is required, a proximal term $-\frac{\rho_i}{2}\|\vP-\vP^{(l)}\|_2^2$, with $\rho_i>0$, can be added without destroying the lower bound, tightness, or gradient-consistency properties.

Because $u_i$ is defined on $\R_+$, the surrogate subproblem must either impose $\widehat R_i(\vP;\vP^{(l)})\geq0$ or extend $u_i$ to an interval containing every surrogate value. In the former case, if $u_i$ is increasing and concave on the relevant operating region, then $u_i(\widehat R_i(\vP;\vP^{(l)}))$ is concave. Moreover, monotonicity of $u_i$ and the bound $\widehat R_i(\vP;\vP^{(l)})\leq R_i(\vP)$ imply $u_i(\widehat R_i(\vP;\vP^{(l)}))\leq u_i(R_i(\vP))$, so the composed surrogate is also a lower bound of the user utility. For a smooth but nonconcave (e.g., smooth S-shaped) utility, this concave-composition argument does not apply in general; instead, one may construct a maximization-consistent surrogate for the smooth composite utility, as below. Only when the utility is genuinely nonsmooth, e.g., kinked at the reference point, is the smooth/nonsmooth decomposition of Proposition~\ref{prop:nonsmooth_sca} required before applying the smooth SCA construction to the differentiable component.

An equivalent variable-splitting representation is also useful for implementation. For utilities satisfying the smooth regularity assumptions of Section~\ref{sec:sca}, let
\begin{equation}
    y_i=g_{ii}P_i,\qquad z_i=N_0+\sum_{j\neq i}g_{ij}P_j,
\end{equation}
so that $R_i=\log_2(1+y_i/z_i)$. The constraints defining $y_i$ and $z_i$ are affine in $\vP$, and the smooth local composite utility $\psi_i(y_i,z_i)=u_i(\log_2(1+y_i/z_i))$ can be approximated by the following maximization-consistent quadratic surrogate
\begin{equation}\label{eq:yz_surrogate_corrected}
\begin{aligned}
    \widetilde\psi_i(y_i,z_i;y_i^{(l)},z_i^{(l)})
    &=\psi_i(y_i^{(l)},z_i^{(l)})\\
    &\quad +A_i^{(l)}(y_i-y_i^{(l)})
    +B_i^{(l)}(z_i-z_i^{(l)})\\
    &\quad -\frac{\tau_i}{2}(y_i-y_i^{(l)})^2
    -\frac{\theta_i}{2}(z_i-z_i^{(l)})^2,
\end{aligned}
\end{equation}
where $A_i^{(l)}=\partial\psi_i/\partial y_i$ and $B_i^{(l)}=\partial\psi_i/\partial z_i$ are evaluated at $(y_i^{(l)},z_i^{(l)})$, and $\tau_i,\theta_i>0$. To invoke the fixed-surrogate-map result of Theorem~\ref{thm:sca_convergence} directly, these regularization parameters are held fixed across outer iterations (or, more generally, chosen through a single iteration-independent surrogate map with a uniform positive lower bound on the strong-concavity modulus). For a kinked utility, these classical derivatives need not exist at the reference point. In that case, the smooth/nonsmooth decomposition of Proposition~\ref{prop:nonsmooth_sca} is used instead of applying \eqref{eq:yz_surrogate_corrected} directly. For the smooth case, the negative signs in the quadratic terms are essential in the maximization form because they make the surrogate strongly concave in $(\vy,\vz)$. When the direct gains satisfy $g_{ii}>0$, the affine map $\vP\mapsto(\vy,\vz)$ is injective because $\vy=\operatorname{diag}(g_{11},\ldots,g_{NN})\vP$. Hence, after eliminating the affine auxiliary variables, the induced surrogate is strongly concave in $\vP$ on the simplex. Provided the resulting smooth composite objective satisfies the regularity assumptions of Theorem~\ref{thm:sca_convergence}, value and gradient consistency together with a step-size rule satisfying that theorem imply that the resulting SCA iteration inherits its stationary-point guarantee.

\begin{remark}[SCA versus SQP]\label{rem:sca_sqp}
Sequential quadratic programming (SQP) is another standard tool for nonconvex constrained optimization \cite{numerical_optimization,SQP_survey}. SQP is often highly effective for small- and medium-scale smooth problems, whereas SCA is attractive here because the surrogate can be tailored to preserve convexity, exploit separability, and produce feasibility-preserving surrogate solutions through convex majorization of nonconvex constraints. The comparison should not be interpreted as a universal dominance claim: practical SQP implementations may include line searches, trust regions, or feasibility restoration, while SCA performance depends on initialization, regularization, and step-size selection. The pilot comparisons in \cite{vaidanis2025SPAWC} are therefore best viewed as evidence that the proposed SCA structure scales favorably for the considered CPT-based allocation instances, rather than as a general statement about all SQP methods.
\end{remark}

\begin{remark}[Computational cost of SCA]\label{rem:sca_complexity}
The computational cost of SCA is surrogate- and inner-solver-dependent. Consequently, the stationarity result in Theorem~\ref{thm:sca_convergence} does not imply a single problem-independent wall-clock or iteration-complexity bound. If $T_{\rm SCA}$ outer iterations are performed, the total arithmetic cost is the sum of the costs of forming and solving the $T_{\rm SCA}$ surrogate subproblems. For the separable simplex subproblem in \eqref{eq:separable_sca_subproblem}, one projected-dual iteration requires $N$ independent scalar primal maximizations together with an $O(N)$ aggregation for the dual update. The overall cost of such an iteration therefore also depends on the cost or numerical accuracy required to solve each scalar maximization. In the scenario-based EUT formulation, each outer iteration additionally evaluates and aggregates the $K$ scenario contributions. CPT uses the same scenario evaluations and, when the scenario ordering is
updated explicitly, additionally sorts the $K$ scenario outcomes and updates their cumulative decision weights, which adds an $O(K\log K)$ ordering step and an $O(K)$ weight-update step per outer iteration. For the linearly coupled ADMM implementation with a dense scenario matrix in $\R^{K\times N}$, the matrix-vector products are $O(KN)$ per inner iteration, in addition to the costs of the corresponding proximal or constrained subproblems. Hence, the end-to-end runtime depends on the number of outer SCA iterations, the required inner-solver accuracy and iterations, and the exploitable problem structure. The pilot execution-time comparisons in \cite{vaidanis2025SPAWC} provide empirical evidence for the considered CPT-based instances, whereas Theorem~\ref{thm:sca_convergence} and the general SCA framework in \cite{SCA_Scutari_Theory_Part1} provide stationarity guarantees rather than a universal runtime bound. Related non-asymptotic analyses of parallel successive convex approximation and block-coordinate updates are available in \cite{RazaviyaynPSCA2014}. For the smooth fixed-step specialization, Corollary~\ref{cor:sca_finite_time} additionally gives an $O(1/\epsilon^2)$ outer-iteration bound in the best-response residual. This bound does not include the cost of solving each inner surrogate problem.
\end{remark}

\section{Deterministic Resource Allocation}\label{sec:deterministic}

This section specializes the general framework to deterministic allocation. The deterministic case is important because it isolates the role of the user-side utility transformation from the additional ranking and probability-weighting effects that arise in stochastic settings.

With the SCA machinery in place, we first consider deterministic systems, where the effect of the utility transformation can be characterized without probability weighting or scenario ranking.

\subsection{Generic Orthogonal Allocation}
Consider the deterministic orthogonal allocation problem
\begin{equation}\label{eq:deterministic_orthogonal}
\begin{aligned}
    \max_{x_1,\ldots,x_N}\quad & \sum_{i=1}^{N} w_i u_i(R_i(x_i))\\
    \text{s.t.}\quad & x_i\geq0,\quad i=1,\ldots,N,\\
    & \sum_{i=1}^{N}x_i\leq X_{\rm tot}.
\end{aligned}
\end{equation}
Here, $R_i(x_i)$ denotes the deterministic operational reward of user $i$, $u_i$ maps this reward into user-side utilization, and $w_i$ is a prioritization coefficient. The problem reduces to classical resource allocation when $u_i(r)=r$; otherwise, the marginal value of the resource depends on both the operational reward function and the shape of the utility function. Classical objectives such as weighted sum-rate and proportional fairness correspond to particular choices of the utility map, whereas more general utility functions modify the effective marginal reward and can therefore alter both the active-user set and the resulting allocation profile. The following result characterizes this effect through the KKT conditions for the generic orthogonal allocation problem.

\begin{theorem}[Generalized water-filling KKT conditions]\label{thm:generalized_water_filling}
Assume $X_{\rm tot}>0$, $w_i\geq0$ for all $i$ and $\sum_i w_i>0$, each $R_i$ is increasing and continuously differentiable, and each $u_i$ is increasing and continuously differentiable. Any local maximizer $\vx^\star$ of \eqref{eq:deterministic_orthogonal} satisfies
\begin{equation}\label{eq:generalized_wf_active}
    w_i u_i'(R_i(x_i^{\star}))R_i'(x_i^{\star})=\mu,\quad x_i^{\star}>0,
\end{equation}
and
\begin{equation}\label{eq:generalized_wf_inactive}
    w_i u_i'(R_i(0))R_i'(0)\leq\mu,\quad x_i^{\star}=0,
\end{equation}
for some resource multiplier $\mu\geq0$, with derivatives at zero interpreted as right derivatives. If $\sum_i x_i^\star<X_{\rm tot}$, then $\mu=0$. If, in addition, every composite function $\phi_i(x_i)=w_i u_i(R_i(x_i))$ is concave on $[0,X_{\rm tot}]$, these KKT conditions are necessary and sufficient for global optimality.
\end{theorem}
\begin{proof}
Set
\begin{equation*}
\phi_i(x_i)\triangleq w_i u_i(R_i(x_i)),
\qquad
\Phi(\vx)\triangleq\sum_{i=1}^{N}\phi_i(x_i).
\end{equation*}
Write the constraints as
\begin{equation*}
g_0(\vx)=\sum_{i=1}^{N}x_i-X_{\rm tot}\leq0,
\qquad
g_i(\vx)=-x_i\leq0.
\end{equation*}
Because these constraints are affine and $X_{\rm tot}>0$, the Mangasarian--Fromovitz constraint qualification holds at every feasible point. Therefore every local maximizer $\vx^\star$ admits multipliers
$\mu\geq0$ and $\nu_i\geq0$ satisfying the KKT system for
\begin{equation*}
\mathcal L(\vx,\mu,\vnu)
=\Phi(\vx)-\mu\left(\sum_{i=1}^{N}x_i-X_{\rm tot}\right)
+\sum_{i=1}^{N}\nu_i x_i.
\end{equation*}
Stationarity and complementary slackness are
\begin{align}
\phi_i'(x_i^\star)-\mu+\nu_i&=0,
\qquad i=1,\ldots,N,
\label{eq:wf_kkt_stationarity}\\
\nu_i x_i^\star&=0,
\qquad i=1,\ldots,N,
\label{eq:wf_kkt_nonnegative_comp}\\
\mu\left(\sum_{i=1}^{N}x_i^\star-X_{\rm tot}\right)&=0.
\label{eq:wf_kkt_budget_comp}
\end{align}
By the chain rule,
\begin{equation*}
\phi_i'(x_i)=w_i u_i'(R_i(x_i))R_i'(x_i).
\end{equation*}
If $x_i^\star>0$, then \eqref{eq:wf_kkt_nonnegative_comp} gives
$\nu_i=0$, and \eqref{eq:wf_kkt_stationarity} yields
\eqref{eq:generalized_wf_active}. If $x_i^\star=0$, then
\begin{equation*}
\nu_i=\mu-\phi_i'(0)\geq0,
\end{equation*}
which is exactly \eqref{eq:generalized_wf_inactive}. If
$\sum_i x_i^\star<X_{\rm tot}$, then
\eqref{eq:wf_kkt_budget_comp} gives $\mu=0$.

Assume now that every $\phi_i$ is concave. For any feasible $\vx$, first-order concavity gives
\begin{align*}
\Phi(\vx)
&\leq \Phi(\vx^\star)
+\sum_{i=1}^{N}\phi_i'(x_i^\star)(x_i-x_i^\star)\\
&=\Phi(\vx^\star)
+\mu\left(\sum_i x_i-\sum_i x_i^\star\right)
-\sum_i\nu_i(x_i-x_i^\star).
\end{align*}
Using $\nu_i x_i^\star=0$, $x_i\geq0$, and
\eqref{eq:wf_kkt_budget_comp}, we obtain
\begin{align*}
\mu\left(\sum_i x_i-\sum_i x_i^\star\right)&\leq0,\\
-\sum_i\nu_i(x_i-x_i^\star)&=-\sum_i\nu_i x_i\leq0.
\end{align*}
Consequently,
\begin{equation*}
\Phi(\vx)\leq\Phi(\vx^\star),
\qquad \forall\vx\in\calX,
\end{equation*}
so the KKT conditions are sufficient for global optimality.
\end{proof}

\begin{corollary}[Unique generalized water-filling response]\label{cor:generalized_wf_map}
Suppose, in addition to the assumptions of Theorem~\ref{thm:generalized_water_filling}, that every composite objective $\phi_i(x)=w_i u_i(R_i(x))$ is strictly concave on $[0,X_{\rm tot}]$. Then \eqref{eq:deterministic_orthogonal} has a unique global maximizer. If the marginal map
\begin{equation*}
m_i(x)\triangleq\phi_i'(x)=w_i u_i'(R_i(x))R_i'(x)
\end{equation*}
is continuous and strictly decreasing, the per-user response to a resource price $\mu\geq0$ is
\begin{equation}\label{eq:generalized_wf_response}
x_i(\mu)=
\begin{cases}
0, & m_i(0)\leq\mu,\\
m_i^{-1}(\mu), & m_i(X_{\rm tot})<\mu<m_i(0),\\
X_{\rm tot}, & \mu\leq m_i(X_{\rm tot}).
\end{cases}
\end{equation}
The unique optimizer satisfies $x_i^\star=x_i(\mu^\star)$ together with $\mu^\star(\sum_i x_i^\star-X_{\rm tot})=0$. 
Under the inherited monotonicity assumptions and strict concavity of every $\phi_i$, the aggregate budget binds; hence a KKT multiplier $\mu^\star$ can be chosen so that $\sum_i x_i(\mu^\star)=X_{\rm tot}$.
\end{corollary}
\begin{proof}
Strict concavity of the separable objective on the convex feasible set gives uniqueness. For a fixed $\mu$, maximizing $\phi_i(x)-\mu x$ on $[0,X_{\rm tot}]$ is a one-dimensional strictly concave problem. Its first-order and endpoint conditions give \eqref{eq:generalized_wf_response}. Because every $\phi_i$ is increasing and strictly concave, a feasible point with $\sum_i x_i<X_{\rm tot}$ cannot be optimal: at least one component can be increased while strictly increasing the objective. Hence the budget binds. The final statement follows from the KKT conditions in Theorem~\ref{thm:generalized_water_filling}.
\end{proof}

\begin{corollary}[Composite-concavity test and Shannon specialization]\label{cor:shannon_composite_concavity}
Let $w_i>0$, and suppose $u_i$ and $R_i$ are twice differentiable on the relevant branch, with $R_i'(x)>0$. For $\phi_i(x)=w_i u_i(R_i(x))$,
\begin{equation}\label{eq:composite_second_derivative}
\phi_i''(x)=w_i\left[u_i''(R_i(x))(R_i'(x))^2+u_i'(R_i(x))R_i''(x)\right].
\end{equation}
Hence $\phi_i$ is concave on an interval if and only if the bracketed quantity in \eqref{eq:composite_second_derivative} is nonpositive throughout that interval. For the Shannon reward
\begin{equation*}
R_i(P_i)=\log_2\!\left(1+\frac{g_iP_i}{N_0}\right),\qquad g_i>0,
\end{equation*}
this condition reduces to
\begin{equation}\label{eq:shannon_utility_concavity_condition}
u_i''(r)\leq (\ln 2)\,u_i'(r)
\end{equation}
for every reward value $r$ attained on the interval. Therefore the user-side utility may be locally convex in the deficit region and the composite allocation objective may nevertheless remain concave, provided its convexity is not stronger than \eqref{eq:shannon_utility_concavity_condition}. 
For a continuous piecewise-$C^2$ utility with finite one-sided derivatives and a downward slope jump at $r_i^0$, and with $R_i$ continuously differentiable and strictly increasing across any point at which $R_i(x)=r_i^0$, the same branchwise condition together with $D^-u_i(r_i^0)\geq D^+u_i(r_i^0)$ is sufficient for concavity of the composite term across the reference point.
\end{corollary}
\begin{proof}
The identity \eqref{eq:composite_second_derivative} follows from the chain rule. For the Shannon reward,
\begin{equation*}
-\frac{R_i''(P_i)}{(R_i'(P_i))^2}=\ln 2,
\end{equation*}
so $\phi_i''(P_i)\leq0$ is equivalent to \eqref{eq:shannon_utility_concavity_condition}. 
At a crossing point $x_0$ with $R_i(x_0)=r_i^0$, continuity of $R_i'$ gives
\begin{equation*}
D^-\phi_i(x_0)=w_i D^-u_i(r_i^0)R_i'(x_0),
\qquad
D^+\phi_i(x_0)=w_i D^+u_i(r_i^0)R_i'(x_0).
\end{equation*}
Since $w_i>0$ and $R_i'(x_0)>0$, the condition
$D^-u_i(r_i^0)\geq D^+u_i(r_i^0)$ implies
$D^-\phi_i(x_0)\geq D^+\phi_i(x_0)$. Together with concavity of the two smooth branches, this preserves concavity across the kink.
\end{proof}

For the Shannon rate $R_i(P_i)=\log_2(1+g_iP_i/N_0)$, \eqref{eq:generalized_wf_active} becomes
\begin{equation}\label{eq:utility_weighted_marginal_rate}
    w_i u_i'(R_i(P_i^{\star}))\frac{g_i}{(N_0+g_iP_i^{\star})\ln 2}=\mu.
\end{equation}
Thus, every differentiable local optimum equalizes utility-weighted marginal rates across active users, while inactive users have no larger marginal value at zero. When the composite objectives are concave, this condition characterizes the global optimum; Corollary~\ref{cor:shannon_composite_concavity} gives a direct test for the Shannon-reward case. With $u_i(r)=r$, the condition reduces to weighted water-filling; the classical unweighted water-filling rule is recovered when, in addition, $w_i=1$ for all users.

Equation~\eqref{eq:utility_weighted_marginal_rate} also explains the active-user patterns observed in the numerical section. A user can be inactive either because its channel marginal rate is small or because its utility marginal value is small at the operating point. Conversely, a user with a moderate channel can remain active when its reference point or deficit sensitivity makes the marginal utility large near the current allocation.

\subsection{Nonorthogonal Power Allocation}
For interference-limited communication, a deterministic utility-aware allocation problem can be written as
\begin{equation}\label{eq:interference_problem}
\begin{aligned}
    \max_{\vP}\quad & \sum_{i=1}^{N} w_i u_i\!\left(\log_2\left(1+\frac{g_{ii}P_i}{\sum_{j\neq i}g_{ij}P_j+N_0}\right)\right)\\
    \text{s.t.}\quad & P_i\geq0,\quad i=1,\ldots,N,\\
    & \sum_{i=1}^{N}P_i\leq P_{\rm tot}.
\end{aligned}
\end{equation}
For interference-coupled wireless systems, weighted sum-rate maximization and related utility formulations are often treated through WMMSE-type transformations and iterative algorithms \cite{ShiWMMSE2011}. For the more general reference-dependent utilities considered here, the rate decomposition in Proposition~\ref{prop:dc_rate} gives a tractable SCA route. This formulation is an interference-channel model. If a downlink NOMA interpretation is intended, then the signal model should be replaced by a SIC-aware expression with a decoding order, a residual interference model if applicable, and rate constraints ensuring decodability.

Using the auxiliary variables introduced in \eqref{eq:yz_surrogate_corrected}, and under the standard assumptions $g_{ij}\geq0$ and $N_0>0$ so that $y_i\geq0$ and $z_i>0$, the same problem can be equivalently written as
\begin{equation}\label{eq:interference_problem_yz}
\begin{aligned}
    \max_{\vP,\vy,\vz}\quad
    &\sum_{i=1}^{N}w_i u_i\!\left(\log_2\left(1+\frac{y_i}{z_i}\right)\right)\\
    \text{s.t.}\quad
    &P_i\geq0,\quad \sum_{i=1}^{N}P_i\leq P_{\rm tot},\\
    &y_i-g_{ii}P_i=0,\quad i=1,\ldots,N,\\
    &z_i-N_0-\sum_{j\neq i}g_{ij}P_j=0,\quad i=1,\ldots,N.
\end{aligned}
\end{equation}
The equality constraints are affine and therefore do not destroy the convexity of the feasible set. For utilities satisfying the smooth regularity assumptions of Theorem~\ref{thm:sca_convergence}, at iteration $l$, replacing each composite term by \eqref{eq:yz_surrogate_corrected} yields a concave quadratic objective over affine and simplex constraints.
If, in addition, $w_i>0$ and $g_{ii}>0$ for all $i$, with fixed $\tau_i,\theta_i>0$, then after eliminating $(\vy,\vz)$ the weighted quadratic regularization induces a strongly concave surrogate in $\vP$. If some weights are zero, strong concavity is not guaranteed by $g_{ii}>0$ alone; it can instead be enforced by adding a global proximal term $-\frac{\rho}{2}\|\vP-\vP^{(l)}\|_2^2,\qquad \rho>0$, to the aggregate surrogate. Hence, the inner problem can be solved using KKT conditions, Lagrangian relaxation, projected-gradient methods, or standard convex optimization solvers. Under the smooth assumptions, and provided the resulting surrogate uses a single iteration-independent map with a uniform strong-concavity modulus and the step-size and exact inner-solution assumptions of Theorem~\ref{thm:sca_convergence} are enforced, every accumulation point is first-order stationary. For genuinely kinked utilities, the composite construction of Proposition~\ref{prop:nonsmooth_sca} should be used instead; under its assumptions, the corresponding accumulation points are Clarke-stationary. 

The deterministic formulations clarify how utility-weighted marginal rewards shape resource allocation. When the system state is random, the allocation instead induces a prospect over scenario-dependent utilities, which is addressed next.

\section{Stochastic Resource Allocation}\label{sec:stochastic}

\subsection{Scenario-Based Wireless Allocation}
Assume that the random state $\vY$ admits a finite scenario representation
\begin{equation}
    \calY=\{(\vy^{(k)},p_k)\}_{k=1}^{K},\quad p_k\geq0,\quad \sum_{k=1}^{K}p_k=1,
\end{equation}
where each $\vy^{(k)}$ is a joint realization of channel, traffic, or task states. The collected utility in scenario $k$ is given by
\begin{equation}    G_k(\vx)=f\!\left(u_1(R_1(\vx,\vy^{(k)})),\ldots,u_N(R_N(\vx,\vy^{(k)}))\right).
\end{equation}
Under EUT with orchestrator Bernoulli utility $U_{\rm orch}$, the stochastic allocation problem is given by
\begin{equation}\label{eq:EUT_stochastic_allocation}
\max_{\vx\in\calX}\; \sum_{k=1}^{K}p_k U_{\rm orch}(G_k(\vx)).
\end{equation}

\begin{proposition}[Scenario SCA for expected utility]\label{prop:scenario_sca}
Assume $\calX$ is nonempty, compact, and convex, the objective in \eqref{eq:EUT_stochastic_allocation} is continuously differentiable with a Lipschitz-continuous gradient on an open set containing $\calX$. For each scenario $k$, assume a single, iteration-independent surrogate map $\widetilde f_k(\vx;\vy)$ is used throughout the run, is jointly continuous in $(\vx,\vy)$, is continuously differentiable in its first argument, has jointly continuous $\nabla_{\vx}\widetilde f_k(\vx;\vy)$, is value- and gradient-consistent with $f_k$, and is uniformly strongly concave in $\vx$ with modulus $\sigma_k\geq\underline{\sigma}>0$ independent of $\vy$. Assume also that every aggregate surrogate problem is solved exactly and that the step sizes satisfy the conditions of Theorem~\ref{thm:sca_convergence}. Then every accumulation point of Algorithm~\ref{alg:SCA} is a first-order stationary point of \eqref{eq:EUT_stochastic_allocation}.
\end{proposition}

\begin{proof}
Let
\begin{align*}
f_k(\vx)&\triangleq U_{\rm orch}(G_k(\vx)),\\
F_{\rm EUT}(\vx)&\triangleq\sum_{k=1}^{K}p_k f_k(\vx),\\
\widetilde F_{\rm EUT}(\vx;\vy)&\triangleq
\sum_{k=1}^{K}p_k\widetilde f_k(\vx;\vy).
\end{align*}
If each $\widetilde f_k(\cdot;\vy)$ is strongly concave with modulus
$\sigma_k\geq\underline\sigma>0$, then, for
$\theta\in[0,1]$,
\begin{align*}
&\widetilde F_{\rm EUT}
\bigl(\theta\vx+(1-\theta)\vz;\vy\bigr)\\
&\quad\geq \theta\widetilde F_{\rm EUT}(\vx;\vy)\\
&\qquad +(1-\theta)\widetilde F_{\rm EUT}(\vz;\vy)\\
&\qquad +\frac{\theta(1-\theta)}{2}
\|\vx-\vz\|_2^2
\sum_{k=1}^{K}p_k\sigma_k.
\end{align*}
Moreover,
\begin{equation*}
\sum_{k=1}^{K}p_k\sigma_k
\geq\underline\sigma\sum_{k=1}^{K}p_k
=\underline\sigma,
\end{equation*}
so the aggregate surrogate is uniformly strongly concave with modulus at least $\underline\sigma$. Moreover,
\begin{align*}
\widetilde F_{\rm EUT}(\vy;\vy)
&=\sum_{k=1}^{K}p_k f_k(\vy)=F_{\rm EUT}(\vy),\\
\nabla_{\vx}\widetilde F_{\rm EUT}(\vy;\vy)
&=\sum_{k=1}^{K}p_k\nabla f_k(\vy)
=\nabla F_{\rm EUT}(\vy).
\end{align*}
Because the sum is finite, joint continuity of $\widetilde F_{\rm EUT}$ and joint continuity of
$\nabla_{\vx}\widetilde F_{\rm EUT}$ are inherited from the scenario surrogates. The aggregate surrogate is also a single, iteration-independent map. Therefore all assumptions of Theorem~\ref{thm:sca_convergence} hold for
$F_{\rm EUT}$ and $\widetilde F_{\rm EUT}$, and every accumulation point $\bar\vx$ satisfies
\begin{equation*}
\nabla F_{\rm EUT}(\bar\vx)^T(\vx-\bar\vx)\leq0,
\qquad \forall\vx\in\calX.
\end{equation*}
\end{proof}

Proposition~\ref{prop:scenario_sca} is a smooth-SCA result. In particular, if $U_{\rm orch}$ is kinked at its reference point and the resulting expected-utility objective is not continuously differentiable, the proposition does not apply directly. The smooth/nonsmooth decomposition and Clarke-stationarity analysis of Proposition~\ref{prop:nonsmooth_sca} should be used instead, provided its assumptions hold.

The EUT formulation treats scenario probabilities linearly. To incorporate reference-dependent evaluation and probability distortion at the orchestrator level, the same scenario prospect can instead be evaluated using CPT.

Under CPT, the scenario outcomes must be sorted relative to the reference point of $U_{\rm orch}$. Let $\pi_k(\vx)$ denote the CPT decision weight assigned to scenario $k$ after sorting the values $G_k(\vx)$ and applying \eqref{eq:CPT_weights}. The CPT allocation problem is given by
\begin{equation}\label{eq:CPT_stochastic_allocation}
    \max_{\vx\in\calX}\; \sum_{k=1}^{K}\pi_k(\vx)U_{\rm orch}(G_k(\vx)).
\end{equation}
The ranking dependence makes \eqref{eq:CPT_stochastic_allocation} piecewise smooth, in contrast to the EUT (Bernoulli utility function), where the scenario probabilities are fixed and independent of the outcome ordering. A practical local SCA implementation fixes the scenario order, the gain/loss partition, and the associated CPT weights at $\vx^{(l)}$, constructs surrogates for the utility terms on that local smooth branch, and updates the ordering and weights at the next outer SCA iteration.

\begin{proposition}[CPT stationarity under eventual branch identification]\label{prop:cpt_local}
Consider the local CPT-SCA iteration for \eqref{eq:CPT_stochastic_allocation}. Suppose there exists an index $L$ and a fixed scenario ordering and gain/loss partition such that every outer iteration $l\geq L$ uses that same CPT branch. Assume that the corresponding smooth branch objective and its SCA surrogates satisfy the assumptions of Theorem~\ref{thm:sca_convergence}. Then every accumulation point of the tail sequence $\{\vx^{(l)}\}_{l\geq L}$ is stationary for that smooth branch. If, at such an accumulation point $\bar\vx$, there are no ties among scenario outcomes and no scenario outcome equals the CPT reference point, then the same branch coincides locally with the original CPT objective, and $\bar\vx$ is locally stationary for \eqref{eq:CPT_stochastic_allocation} on that nondegenerate branch.
\end{proposition}
\begin{proof}
After iteration $L$, the scenario ordering and gain/loss partition are fixed. Hence, the CPT decision weights are fixed constants $\bar\pi_k$, and the tail sequence is precisely a smooth SCA sequence for
\begin{equation*}
F_{\bar\pi}(\vx)\triangleq\sum_{k=1}^{K}\bar\pi_kU_{\rm orch}(G_k(\vx)).
\end{equation*}
Removing finitely many initial iterations does not alter the step-size assumptions of Theorem~\ref{thm:sca_convergence}. Therefore, every accumulation point $\bar\vx$ of the tail satisfies
\begin{equation*}
\nabla F_{\bar\pi}(\bar\vx)^T(\vx-\bar\vx)\leq0,
\qquad \forall\vx\in\calX,
\end{equation*}
or equivalently
\begin{equation*}
0\in-\sum_{k=1}^{K}\bar\pi_k\nabla\!\bigl[U_{\rm orch}(G_k(\bar\vx))\bigr]+N_{\calX}(\bar\vx).
\end{equation*}
If $\bar\vx$ has neither outcome ties nor reference-point crossings, then
\begin{equation*}
\Delta\triangleq\min\left\{\min_{j\neq k}|G_j(\bar\vx)-G_k(\bar\vx)|,\min_k|G_k(\bar\vx)-z_0|\right\}>0.
\end{equation*}
Continuity of the finite family $\{G_k\}_{k=1}^{K}$ then yields a neighborhood of $\bar\vx$ in which neither the ordering nor the gain/loss classification changes. Hence $F_{\bar\pi}$ agrees with the original CPT objective throughout that neighborhood, proving the local branch-stationarity claim.
\end{proof}
We note that Proposition~\ref{prop:cpt_local} requires eventual branch identification. It does not provide a global convergence guarantee at outcome ties, at reference-point crossings, or for trajectories whose scenario ordering changes infinitely often.

\subsection{Linearly Coupled Stochastic Allocation}
A useful special case arises when the scenario rewards are affine in the allocation vector. Let $\vRmat\in\R^{K\times N}$ collect $K$ observed or simulated scenarios, with each row representing an uncertain return, channel-state reward, or task-effectiveness coefficient, and introduce $\vz=\vRmat\vx$. Then, the EUT-based stochastic problem can be written as
\begin{equation}\label{eq:linear_scenario_eut}
\begin{aligned}
\max_{\vx\in\calX,\vz}\quad & \sum_{k=1}^{K}p_k U_{\rm orch}(z_k)\\
    \text{s.t.}\quad & \vz=\vRmat\vx.
\end{aligned}
\end{equation}
This form includes the portfolio example as a noncommunication benchmark. Within the present communication framework, it represents the special case in which each scenario reward is affine in the allocation vector; for example, $\vRmat$ may contain scenario-dependent linearized channel, reliability, or semantic-effectiveness coefficients. It should not be interpreted as a generic nonlinear wireless-rate model. The auxiliary variable $\vz$ separates the scenario-wise utility evaluation from the resource-allocation constraint, which is useful for SCA, ADMM, and projected first-order methods.

For utilities with a reference-point kink, \eqref{eq:orch_decomposition} yields
\begin{equation}
    U_{\rm orch}(z_k)=\widehat U_{\rm orch}(z_k)+\widetilde U_{\rm orch}(z_k),
\end{equation}
where $\widehat U_{\rm orch}$ is smooth and possibly nonconcave, while $\widetilde U_{\rm orch}$ is concave for a downward kink. Provided the corrected component $\widehat U_{\rm orch}$ satisfies the smooth regularity assumptions required in Proposition~\ref{prop:nonsmooth_sca} on the reachable set $\{\vRmat\vx:\vx\in\calX\}$, the SCA surrogate keeps $\widetilde U_{\rm orch}$ exactly and applies the proximal linearization only to the smooth part. This is the stochastic counterpart of Proposition~\ref{prop:nonsmooth_sca} and avoids treating the reference-point derivative discontinuity as a numerical artifact.

\begin{remark}[ADMM solution of linearly coupled SCA subproblems]
In scenario-based formulations, an SCA subproblem may become separable after introducing auxiliary variables for the scenario-wise rewards. For instance, if the local surrogate can be written in the linearly coupled form
\begin{equation}
\begin{aligned}
    \min_{\vx,\vz}\quad & \Phi^{(l)}(\vz)+I_{\calX}(\vx)\\
    \text{subject to}\quad & \vz=\vA\vx,
\end{aligned}
\label{eq:admm_sca_subproblem}
\end{equation}
where $\vz$ collects the scenario-wise linear rewards (or auxiliary variables entering the scenario utilities), $\vA$ is a scenario matrix, $\Phi^{(l)}$ is the convex majorant/minimization form generated at the current SCA iteration, and $I_{\calX}$ is the indicator function of the feasible set, assume that $\Phi^{(l)}$ is proper, closed, and convex, $\calX$ is nonempty, closed, and convex, and the inner problem has a solution. Under these standard two-block convex-ADMM conditions, the inner problem can be solved by ADMM. With the scaled dual variable $\vnu$ and penalty $\rho>0$, the iterations are
\begin{subequations}
\begin{align}
    \vx^{(r+1)}&=\argmin_{\vx\in\calX}
    \frac{\rho}{2}\left\|\vA\vx-\vz^{(r)}+\vnu^{(r)}\right\|_2^2,\\
    \vz^{(r+1)}&=\argmin_{\vz}
    \Phi^{(l)}(\vz)+\frac{\rho}{2}\left\|\vA\vx^{(r+1)}-\vz+\vnu^{(r)}\right\|_2^2,\\
    \vnu^{(r+1)}&=\vnu^{(r)}+\vA\vx^{(r+1)}-\vz^{(r+1)}.
\end{align}
\end{subequations}
If CPT ordering constraints are fixed at the outer SCA iterate and $\Phi^{(l)}$ is separable so that the $\vz$-subproblem reduces to a weighted isotonic-regression problem with chain constraints, the update can be computed efficiently by a pool-adjacent-violators algorithm or a generalized isotonic-regression routine. For a general convex $\Phi^{(l)}$, no such PAVA reduction is implied by \eqref{eq:admm_sca_subproblem} alone. ADMM is used only to solve the convexified inner problem. The outer convergence results established in this paper assume exact solution of each SCA subproblem. Therefore, to invoke those results directly, the ADMM inner loop must recover the exact surrogate optimizer in the limiting analysis. A finite-tolerance ADMM stopping rule requires an additional inexact-SCA argument (for example, with suitably controlled inner errors), which is not established here. For CPT, the branch-stationarity conclusion also remains conditional on the local ordering/reference-point assumptions stated in Proposition~\ref{prop:cpt_local}.
\end{remark}

In summary, the stochastic formulation can be solved at three levels of complexity. EUT with smooth utilities leads to a scenario-average SCA problem. CPT adds local ranking and probability-weighting updates at the outer iteration, with the branch-stationarity qualification of Proposition~\ref{prop:cpt_local}. Linearly coupled scenario models additionally permit ADMM or projected-subgradient implementations for the inner convexified problem.

\section{Simulation Results}\label{sec:simulations}
In this section, we numerically illustrate and validate the analytical results. The first experiment isolates the effect of the core utility through a scalar resource-price response. The second evaluates the SCA solver sensitivity to the proximal regularization parameter using a common fixed diagnostic projected-gradient stationarity measure. The third studies the effects of total power, reference level, and deficit sensitivity in a globally concave orthogonal benchmark. We then compare classical deterministic objectives, illustrate the soft-QoS penalty continuation, verify the generalized water-filling KKT residual, and finish with a fully specified fixed-scenario EUT/CPT comparison. We note that all powers and channel gains below are dimensionless normalized quantities unless stated otherwise.

\subsection{Core Utility Price Responses}
To isolate the effect of the core utility without leaving the other users or the total budget unspecified, we use the scalar resource-price response associated with the dual decomposition of an orthogonal allocation problem:
\begin{equation}\label{eq:core_price_response}
P^\star(g;\mu)\in\argmax_{0\leq P\leq P_{\max}}
\left\{u\!\left(1+\frac{gP}{N_0}\right)-\mu P\right\}.
\end{equation}
We fix $N_0=1$, $P_{\max}=5$, and $\mu=0.22$, and compare
\begin{equation}\label{eq:core_utility_cases}
\begin{aligned}
 u_{\rm exp}(s)&=1-e^{-s},&
 u_{\log}(s)&=\log s,\\
 u_{0.4}(s)&=s^{0.4},&
 u_{0.8}(s)&=s^{0.8},\qquad s\geq1.
\end{aligned}
\end{equation}
Here $s=1+gP/N_0$ is used as a normalized scalar operational metric rather than as a Shannon-rate expression, as the purpose of this experiment is to isolate the effect of utility curvature.

\begin{figure}[!t]
    \centering
    \includegraphics[width=.75\linewidth]{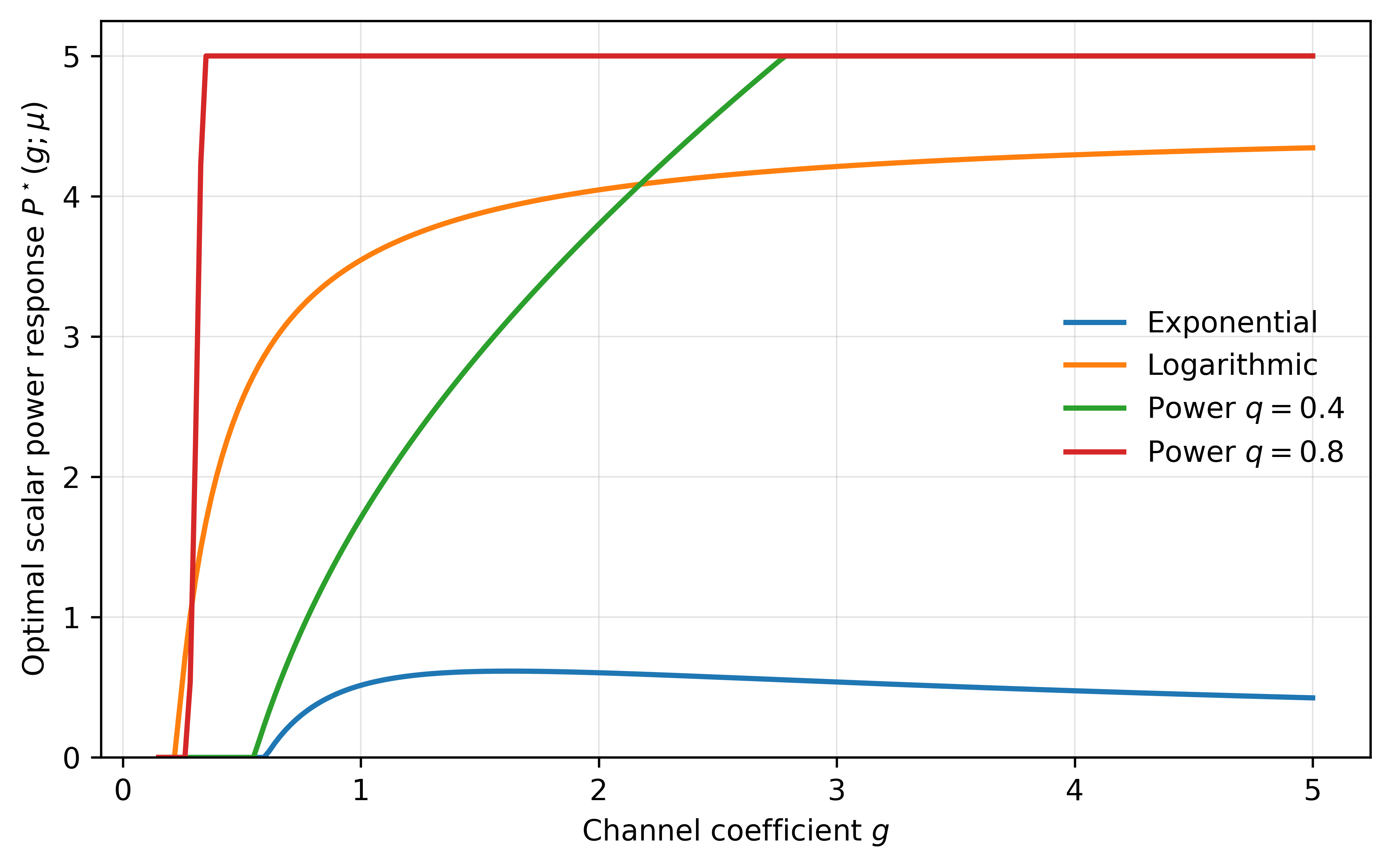}
    \caption{Optimal scalar power response $P^\star(g;\mu)$ for the four core utilities in \eqref{eq:core_utility_cases}, with $N_0=1$, $P_{\max}=5$, and resource price $\mu=0.22$.}
    \label{fig:core_power_policy}
\end{figure}

\begin{figure}[!t]
    \centering
    \includegraphics[width=.75\linewidth]{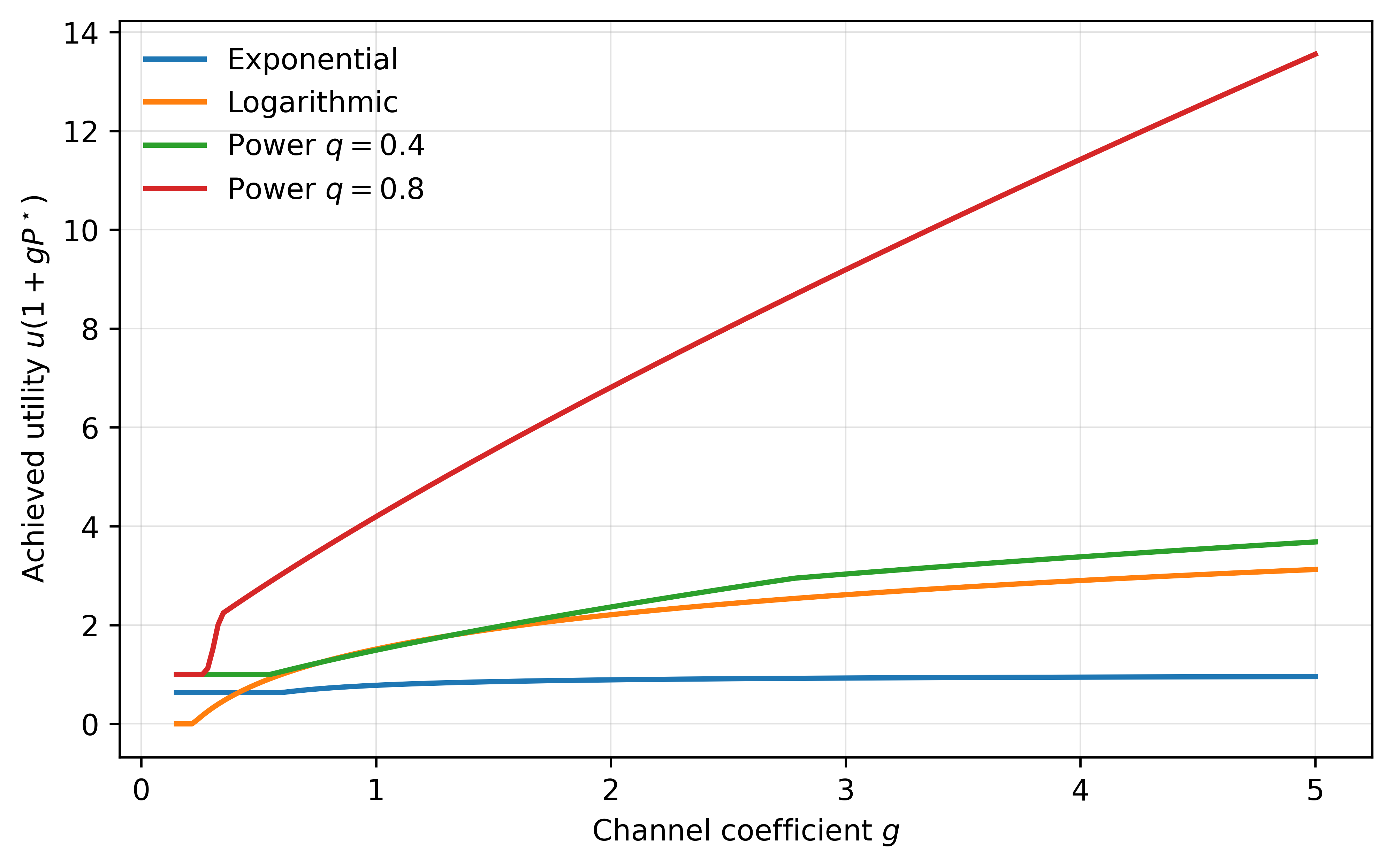}
    \caption{Achieved utility $u(1+gP^\star)$ for the same fixed-price responses as Fig.~\ref{fig:core_power_policy}.}
    \label{fig:core_utility_policy}
\end{figure}

All four functions are increasing and strictly concave on the plotted domain. Consequently, \eqref{eq:core_price_response} is a one-dimensional strictly concave maximization problem for $g>0$. For $g=0$, the unique solution is $P^\star=0$ because $\mu>0$. Hence the scalar response is uniquely determined for all
$g\geq0$ considered here. Figs.~\ref{fig:core_power_policy} and~\ref{fig:core_utility_policy} are fixed-price response curves, showing directly how the marginal-value curve changes the amount of resource purchased at a common price and the utility achieved from that resource.

\subsection{SCA Sensitivity Experiment}
We next focus on \eqref{eq:interference_problem} with equal prioritization weights and instantiate the extended utility in \eqref{eq:Extension_KW} explicitly. We use
\begin{equation}
    g_1(r,r^0)=-(r-r^0), r\geq r^0,
    \qquad
    g_2(r,r^0)=r-r^0, r<r^0,
    \label{eq:tableI_branch_functions}
\end{equation}
so that the gain branch is increasing and concave for $\lambda_1,\alpha,\gamma_1>0$, while the deficit branch is increasing and convex for $\lambda_2<0$, $\beta,\gamma_2>0$, and $\lambda_3\geq0$. The one-sided marginal utilities at the reference point are
\begin{equation}
    D^+u(r^0)=\frac{\lambda_1}{\gamma_1},
    \qquad
    D^-u(r^0)=-\frac{\lambda_2}{\gamma_2}+\lambda_3.
    \label{eq:tableI_reference_slopes}
\end{equation}
The four numerical cases below satisfy $D^-u(r^0)=D^+u(r^0)$ and are therefore slope-matched. Since each branch is smooth with bounded second derivative on the compact operating interval, the derivative is locally Lipschitz across the matched reference point. Moreover, because the interference-rate functions are smooth on the compact power simplex for $N_0>0$, the resulting composite objective $F$ is continuously differentiable with a Lipschitz-continuous gradient on the feasible operating region, as required by the smooth SCA analysis.

For this experiment, $N=5$, $N_0=1$, $P_{\rm tot}=5$, $w_i=1$, and $r_i^0=1$. The direct gains are $[g_{11},\ldots,g_{55}]=[2.0,1.7,1.4,1.1,0.9]$, and $g_{ij}=0.08$ for $i\neq j$. Every run starts from $P_i^{(0)}=1$. In the parameter order $(\lambda_1,\lambda_2,\lambda_3,\alpha,\beta,\gamma_1,\gamma_2)$, the four cases are
\begin{align*}
\vartheta_1&=(1,-0.4,0.6,1,1,1,1),\\
\vartheta_2&=(0.8,-0.6,0.5,2,2,0.8,1.2),\\
\vartheta_3&=(1,-0.8,1,1,1,0.5,0.8),\\
\vartheta_4&=(0.8,-0.8,1,2,2,0.4,0.8).
\end{align*}
Cases 1-2 have reference-point slope one, while Cases 3-4 have slope two. Cases 2 and 4 use stronger curvature and post-reference saturation.

For a fixed $\rho$, the exact best response of \eqref{eq:proximal_surrogate} on the power simplex $\calX=\left\{\vP\geq0:\sum_iP_i\leq P_{\rm tot}\right\}$ is
\begin{equation}\label{eq:tableI_best_response}
\widehat{\vP}_{\rho}^{(l)}
=\Pi_{\calX}\!\left(\vP^{(l)}+\rho^{-1}\nabla F(\vP^{(l)})\right),
\end{equation}
followed by
\begin{equation}\label{eq:tableI_step_rule}
\vP^{(l+1)}=\vP^{(l)}+\gamma_l
\left(\widehat{\vP}_{\rho}^{(l)}-\vP^{(l)}\right),
\qquad \gamma_l=(l+1)^{-0.6}.
\end{equation}
The choice $\gamma_l=(l+1)^{-0.6}$ satisfies $\gamma_l\to0$, $\sum_l\gamma_l=\infty$, and $\sum_l\gamma_l^2<\infty$, as required by
Theorem~\ref{thm:sca_convergence}. Because the algorithmic best-response displacement depends on $\rho$, comparisons across different regularization values use one fixed diagnostic projected-gradient mapping,
\begin{equation}\label{eq:projected_gradient_mapping}
\mathcal G_{\rm ref}(\vP)
\triangleq \bar\rho\left[
\Pi_{\calX}\!\left(\vP+\bar\rho^{-1}\nabla F(\vP)\right)-\vP
\right],\qquad \bar\rho=1.
\end{equation}
The diagnostic parameter $\bar\rho$ is held fixed for every run and is distinct from the algorithmic regularization $\rho$. For any $\bar\rho>0$, $\mathcal G_{\rm ref}(\vP)=0$ if and only if $\vP$ is first-order stationary for the simplex-constrained smooth problem. Thus, all regularization choices are terminated using the same stationarity measure, $\|\mathcal G_{\rm ref}(\vP^{(l)})\|_2\leq10^{-6}$.

\begin{table}[!t]
    \centering
    \caption{SCA sensitivity experiment using the common fixed-diagnostic projected-gradient stopping criterion $\|\mathcal G_{\rm ref}\|_2\leq10^{-6}$. Each entry reports ``final objective (outer iterations).'' Values across rows are not directly comparable because the utility functions differ.}
    \label{tab:sca_output}
    \small
    \begin{tabular}{c|ccc}
        \hline
        Case & $\rho=0.25$ & $\rho=0.50$ & $\rho=1.00$ \\
        \hline
        1 & $0.181789\;(66)$ & $0.181789\;(378)$ & $0.181789\;(2037)$ \\
        2 & $0.124337\;(78)$ & $0.124337\;(290)$ & $0.124337\;(1780)$ \\
        3 & $0.267043\;(50)$ & $0.267043\;(88)$ & $0.267043\;(381)$ \\
        4 & $0.138962\;(91)$ & $0.138962\;(79)$ & $0.138962\;(98)$ \\
        \hline
    \end{tabular}
\end{table}

\begin{figure}[!t]
    \centering
    \includegraphics[width=0.75\linewidth]{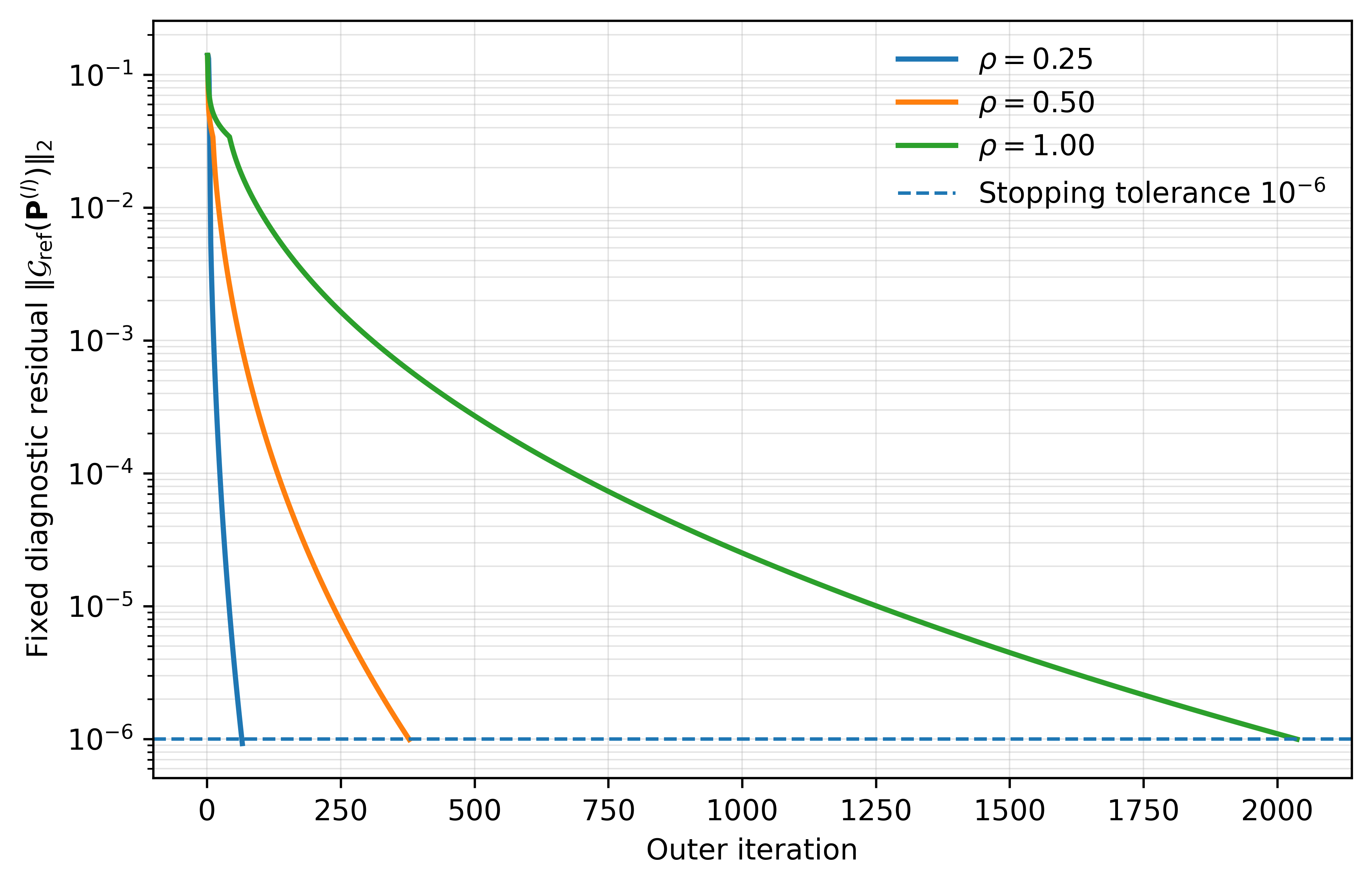}
    \caption{Convergence of the fixed-diagnostic projected-gradient stationarity residual $\|\mathcal G_{\rm ref}(\vP^{(l)})\|_2$ for utility Case~1. All runs use the same equal-power initialization and $\gamma_l=(l+1)^{-0.6}$, and terminate at the common tolerance $10^{-6}$.}
    \label{fig:interference_convergence}
\end{figure}

For every utility case, the three regularization choices converge to the same reported objective value within numerical precision from the common initialization, while the iteration count changes substantially. Figure~\ref{fig:interference_convergence} shows Case~1 using the common residual in \eqref{eq:projected_gradient_mapping}: the runs terminate after $66$, $378$, and $2037$ outer iterations for $\rho=0.25$, $0.50$, and $1.00$, respectively. This is a fixed-instance solver-sensitivity result from the common initialization, not a global-optimality claim or a universal ordering of regularization parameters.

\subsection{Reference, Deficit-Sensitivity, and Power Sweeps}
\begin{figure*}[!t]
    \centering
    \setkeys{Gin}{width=0.31\linewidth}
    \subfloat[Total-power sweep]{\includegraphics{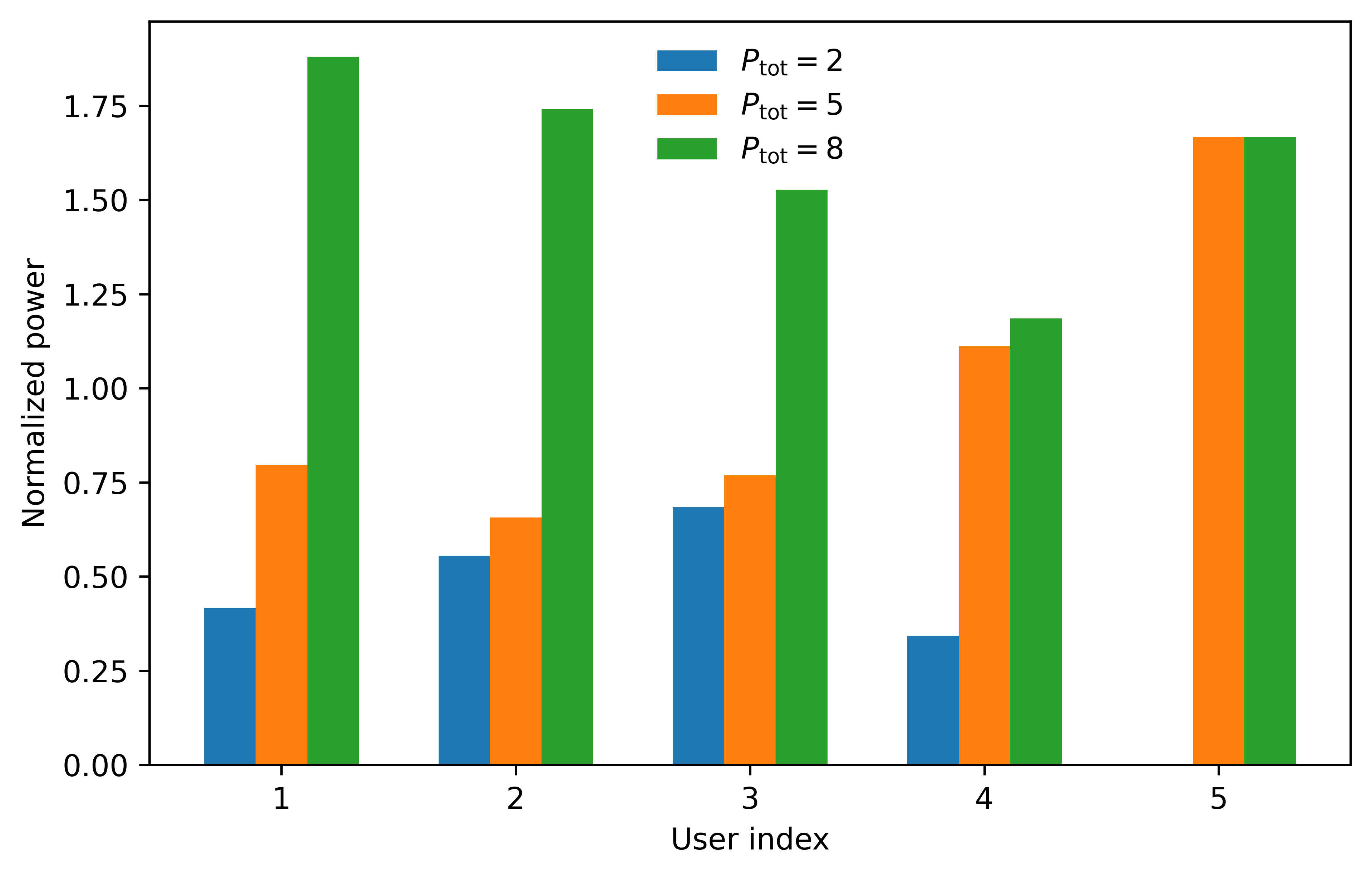}}\hfil
    \subfloat[Reference-level sweep]{\includegraphics{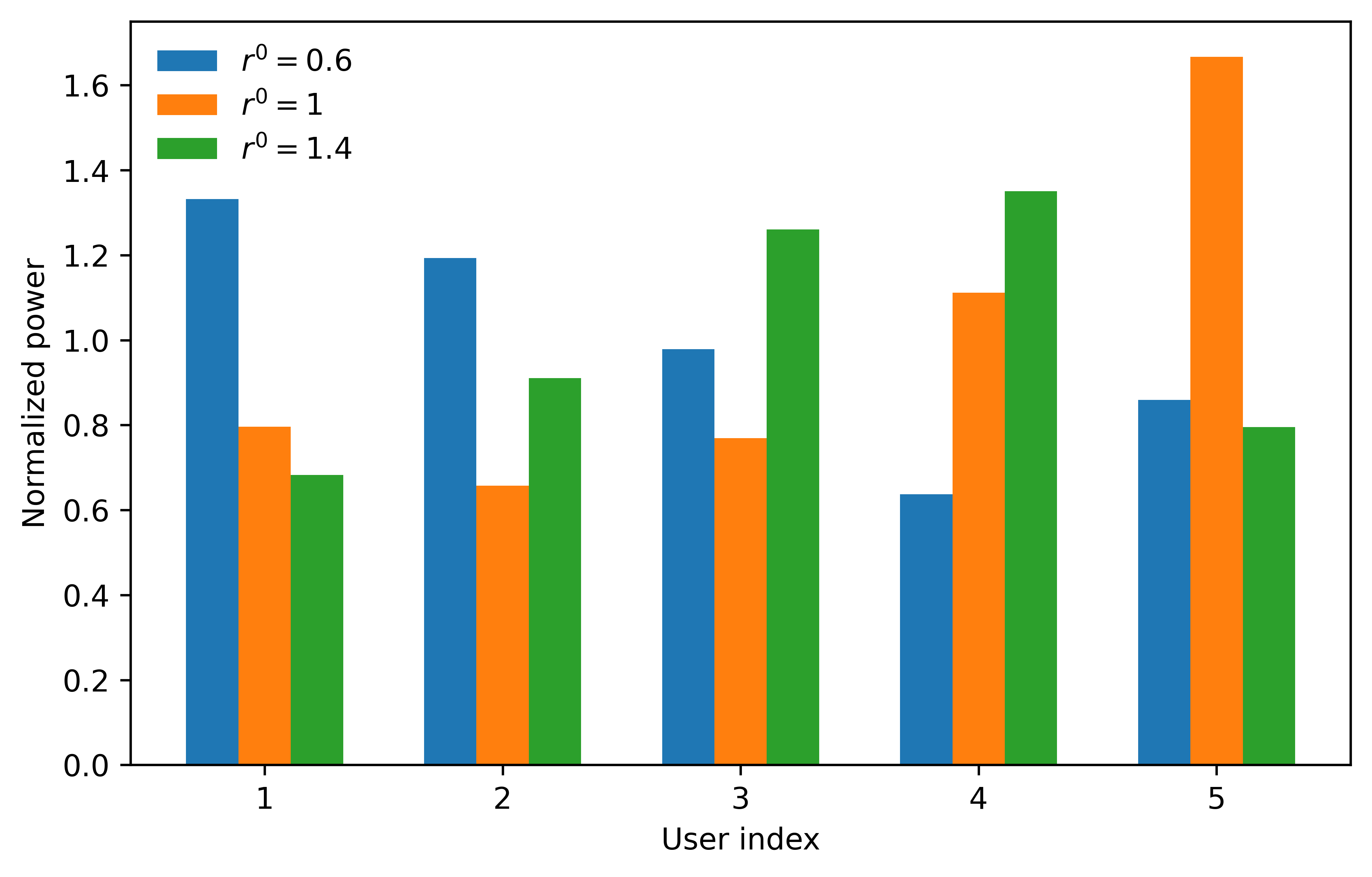}}\hfil
    \subfloat[Deficit-sensitivity sweep]{\includegraphics{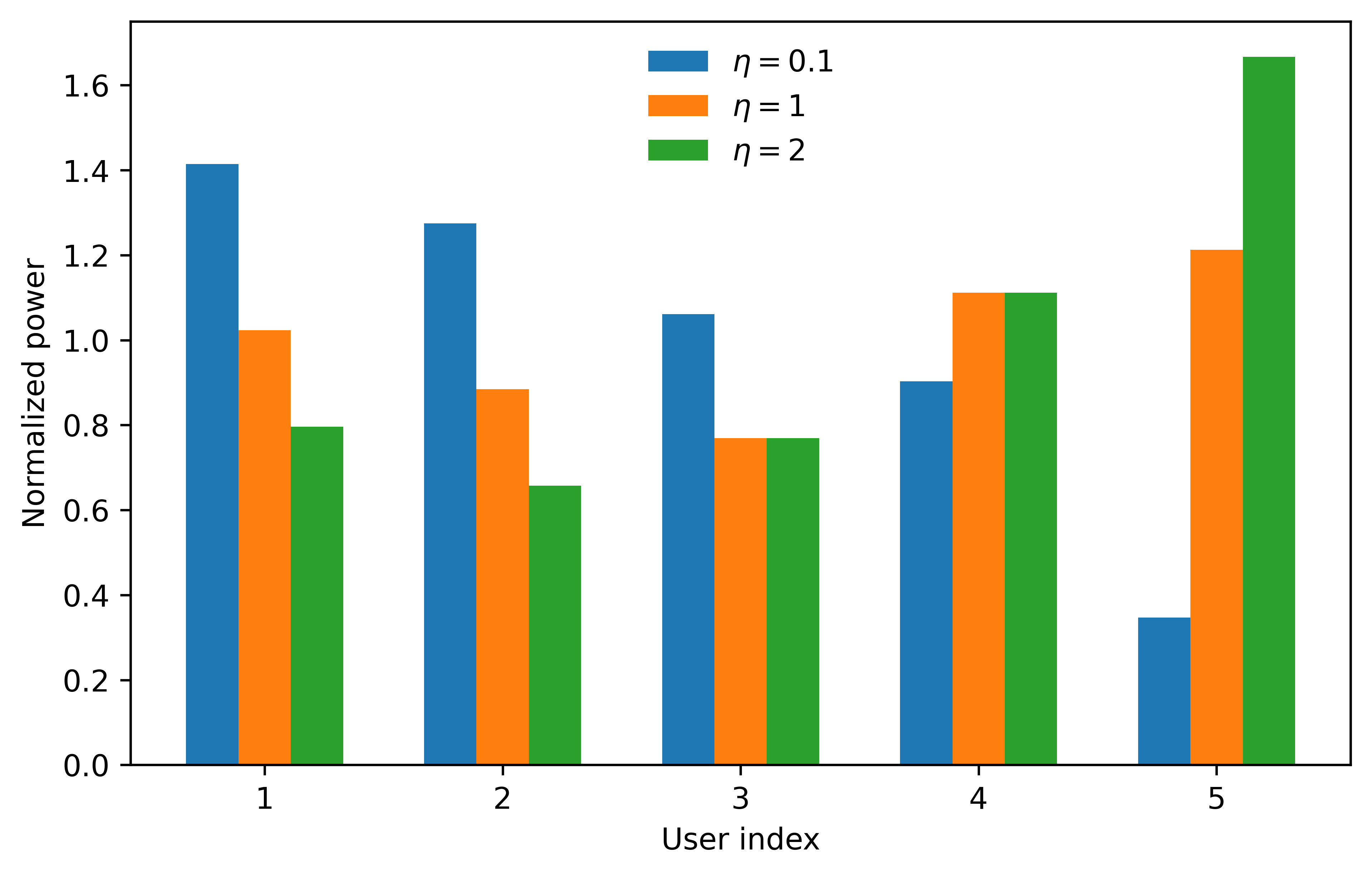}}
    \caption{Globally optimal orthogonal allocations for the utility in \eqref{eq:soft_utility}. All axes show dimensionless power units.}
    \label{fig:sensitivity}
\end{figure*}

To study separately the effects of the reference level, deficit
sensitivity, and total resource budget, we use the explicitly normalized
reference-dependent utility 
\begin{equation}\label{eq:soft_utility}
u_i(r)=r-r_i^0-\eta\,(r_i^0-r)_+,\qquad \eta>0.
\end{equation}
Its slope is $1+\eta$ below the reference and $1$ above it, so it is nondecreasing and concave with a downward kink at $r_i^0$. With orthogonal Shannon rewards $R_i(P_i)=\log_2(1+g_iP_i)$, each composite $u_i(R_i(P_i))$ is strictly concave on the feasible power interval. Hence the aggregate objective is strictly concave on
the power simplex, and each instance in Fig.~\ref{fig:sensitivity} has a
unique global optimum rather than merely a stationary solution.

We fix $N=5$, $N_0=1$, and channel gains $[2.4,1.8,1.3,0.9,0.6]$. Panel (a) varies $P_{\rm tot}\in\{2,5,8\}$ with common $r^0=1$ and $\eta=2$. Panel (b) varies the common reference $r^0\in\{0.6,1,1.4\}$ with $P_{\rm tot}=5$ and $\eta=2$. Panel (c) varies $\eta\in\{0.1,1,2\}$ with $P_{\rm tot}=5$ and $r^0=1$. The results show, for this fixed channel vector, that changing the reference point or deficit sensitivity alters the users' marginal utility of power and can therefore redirect power toward users operating below their reference levels. Increasing the total budget also changes which users can be moved through their reference regions. These are instance-specific comparative statics, not universal user-ordering claims.

\subsection{Benchmark Comparison with Classical Objectives}
We next use $N=8$, $N_0=1$, $P_{\rm tot}=8$, deterministic channel gains $[3.2,2.5,1.9,1.45,1.10, \newline 0.82,0.62,0.45]$, and reference rates $[1.75,1.55,1.35,1.15,0.95,0.85,0.75,0.65]$. The reference-dependent policy uses \eqref{eq:soft_utility} with $\eta=1$. Its objective is therefore
\begin{equation}\label{eq:deterministic_soft_objective}
\sum_{i=1}^{N}\left[R_i(P_i)-r_i^0-(r_i^0-R_i(P_i))_+\right].
\end{equation}
Equal power uses $P_i=P_{\rm tot}/N$. WSR maximizes $\sum_iR_i$; PF maximizes $\sum_i\log R_i$; the $\alpha$-fair baseline with $\alpha=2$ maximizes $-\sum_iR_i^{-1}$; and hard-QoS WSR adds $R_i\geq r_i^0$. For the orthogonal Shannon model used here, the optimized objectives are
concave on their respective effective domains. In particular, the PF and $\alpha$-fair objectives are defined only for strictly positive rates, so their feasible domains exclude allocations with $R_i=0$.

\begin{table}[t]
\caption{Deterministic baseline comparison for one fixed normalized channel instance. ``Soft util.'' is \eqref{eq:deterministic_soft_objective}, ``Viol.'' counts users with $R_i<r_i^0$, and ``Jain'' is computed over achieved rates.}
\label{tab:deterministic_baselines}
\centering
\begin{tabular}{lccccc}
\hline
Policy & Sum rate & Soft util. & Min rate & Viol. & Jain \\
\hline
Equal power & 9.873 & 0.705 & 0.536 & 2 & 0.855 \\
WSR & 10.383 & 0.088 & 0.000 & 3 & 0.680 \\
PF & 9.634 & 0.601 & 0.617 & 1 & 0.897 \\
$\alpha$-fair & 9.158 & -0.199 & 0.766 & 3 & 0.951 \\
Hard-QoS WSR & 9.740 & 0.740 & 0.650 & 0 & 0.853 \\
Ref. dep. utility & 9.963 & 0.782 & 0.469 & 1 & 0.817 \\
\hline
\end{tabular}
\end{table}

\begin{figure}[t]
    \centering
    \includegraphics[width=0.75\linewidth]{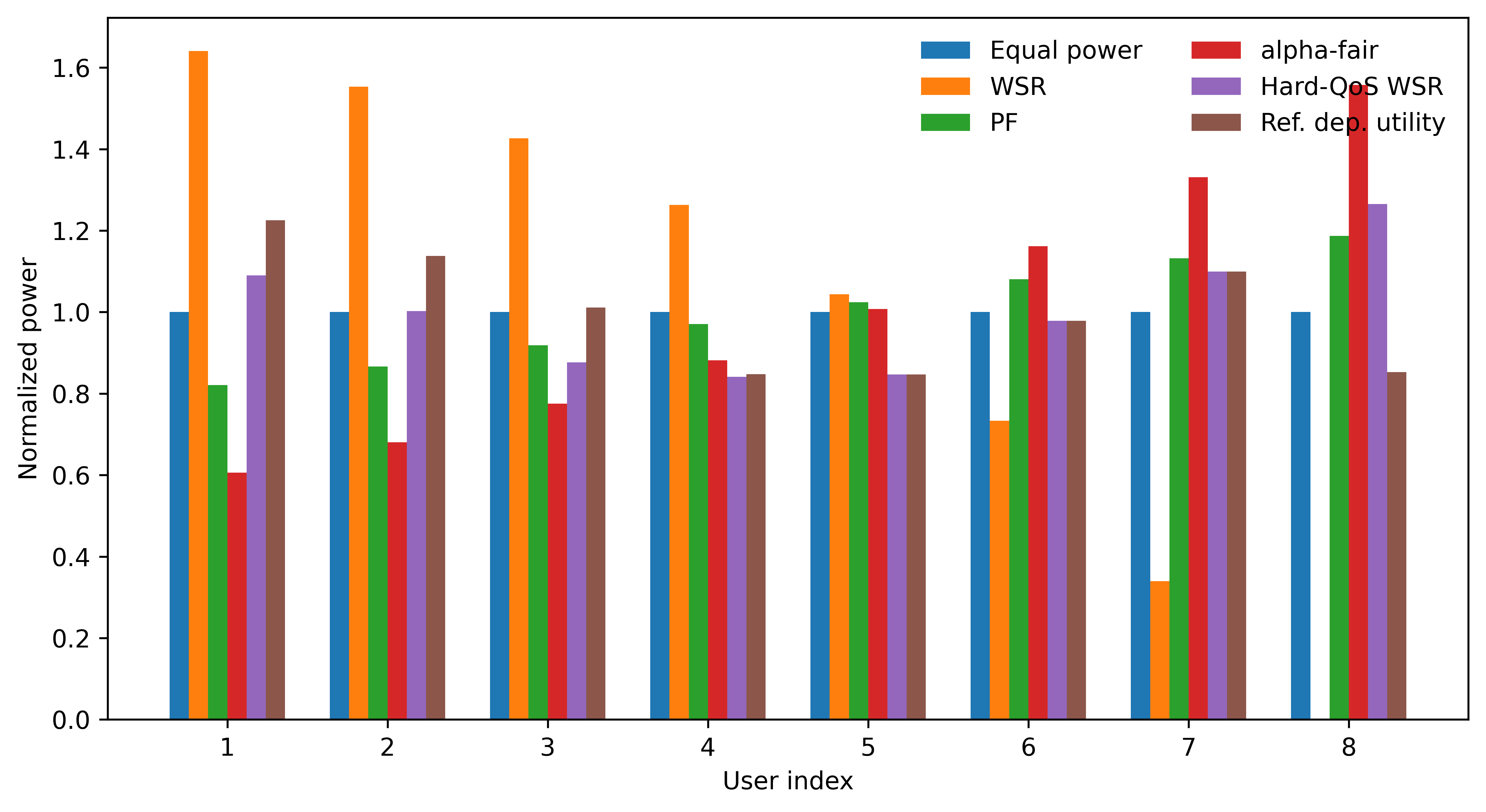}
    \caption{Normalized power allocations for the deterministic policies in Table~\ref{tab:deterministic_baselines}.}
    \label{fig:deterministic_allocations}
\end{figure}

Table~\ref{tab:deterministic_baselines} and Fig.~\ref{fig:deterministic_allocations} make the soft-versus-hard distinction explicit. WSR attains the largest sum rate but violates three reference-rate targets. Hard-QoS WSR enforces all targets and sacrifices sum rate. The reference-dependent policy attains the largest value of the common soft utility metric in \eqref{eq:deterministic_soft_objective}, as expected from its optimization criterion, and reduces the number of violated reference-rate targets relative to WSR, while still leaving one user below its target.

\begin{figure}[t]
    \centering
    \includegraphics[width=.75\linewidth]{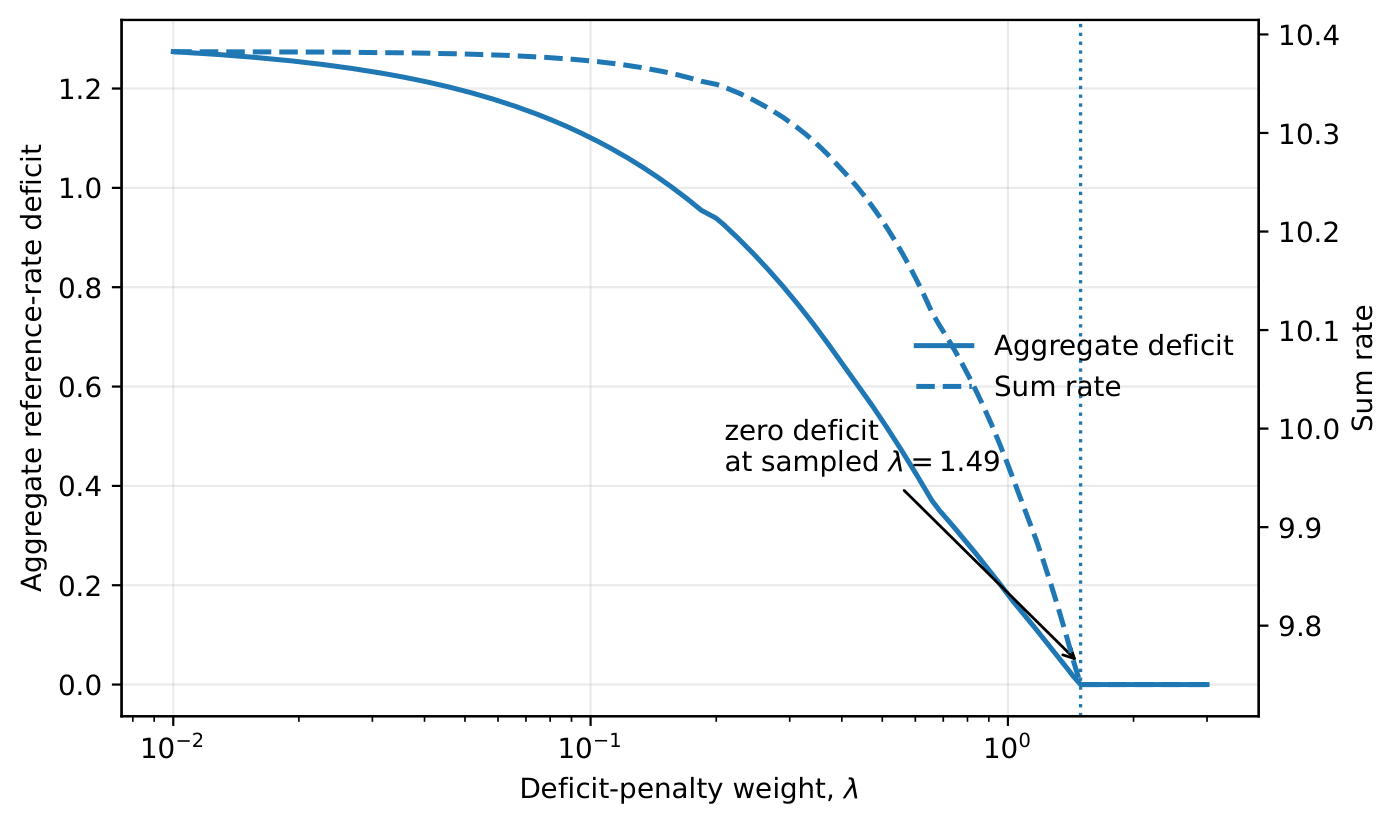}
    \caption{Globally solved soft-QoS penalty continuation on the deterministic benchmark: aggregate reference-rate deficit and sum rate versus $\lambda$.}
    \label{fig:soft_qos_penalty_validation}
\end{figure}

To illustrate Corollary~\ref{cor:soft_qos_rate} separately, we consider the penalty-continuation problem on the same fixed channel and reference vector,
\begin{eqnarray}
\max_{\vP\in\calX}&& S(\vP)-\lambda D(\vP),\\
S(\vP)&=&\sum_{i=1}^{N}R_i(P_i),\\
D(\vP)&=&\sum_{i=1}^{N}\bigl(r_i^0-R_i(P_i)\bigr)_+.
\end{eqnarray}
For this specialization, $r\mapsto r-\lambda(r_i^0-r)_+$ is nondecreasing and concave, so the separable objective is concave and each penalized instance
admits a global optimum. Figure~\ref{fig:soft_qos_penalty_validation} shows that the aggregate deficit decreases with $\lambda$ and becomes zero at approximately $\lambda\simeq1.49$, after which the solution coincides numerically with hard-QoS WSR. The experiment is consistent with the $O(1/\lambda)$ upper bound of Corollary~\ref{cor:soft_qos_rate}; it is not presented as an empirical proof of an exact asymptotic decay rate.

\subsection{Validation of the Generalized Water-Filling Condition}
We next numerically validate the generalized water-filling KKT structure of Theorem~\ref{thm:generalized_water_filling} using a fully specified smooth reference-dependent utility. We use the same deterministic channel gains and reference-rate vector as Table~\ref{tab:deterministic_baselines}, set $P_{\rm tot}=4$, $w_i=1$, and choose the slope-matched utility corresponding to $\vartheta_1$ in the SCA experiment:
\begin{equation}\label{eq:wf_validation_utility}
u_i(r)=
\begin{cases}
1-e^{-(r-r_i^0)}, & r\geq r_i^0,\\
-0.4\bigl(1-e^{r-r_i^0}\bigr)+0.6(r-r_i^0), & r<r_i^0.
\end{cases}
\end{equation}
For the gain branch, $u_i''(r)<0$. For the deficit branch,
$u_i'(r)=0.4e^{r-r_i^0}+0.6$ and $u_i''(r)=0.4e^{r-r_i^0}$, so $\frac{u_i''(r)}{u_i'(r)}\leq0.4<\ln2$. Hence Corollary~\ref{cor:shannon_composite_concavity}
verifies strict concavity of every composite $u_i(R_i(P_i))$ over the feasible interval. The resulting orthogonal problem is therefore strictly concave
on the power simplex and has a unique global optimum, and the KKT
conditions are consequently necessary and sufficient.

\begin{figure}[t]
    \centering
    \includegraphics[width=.75\linewidth]{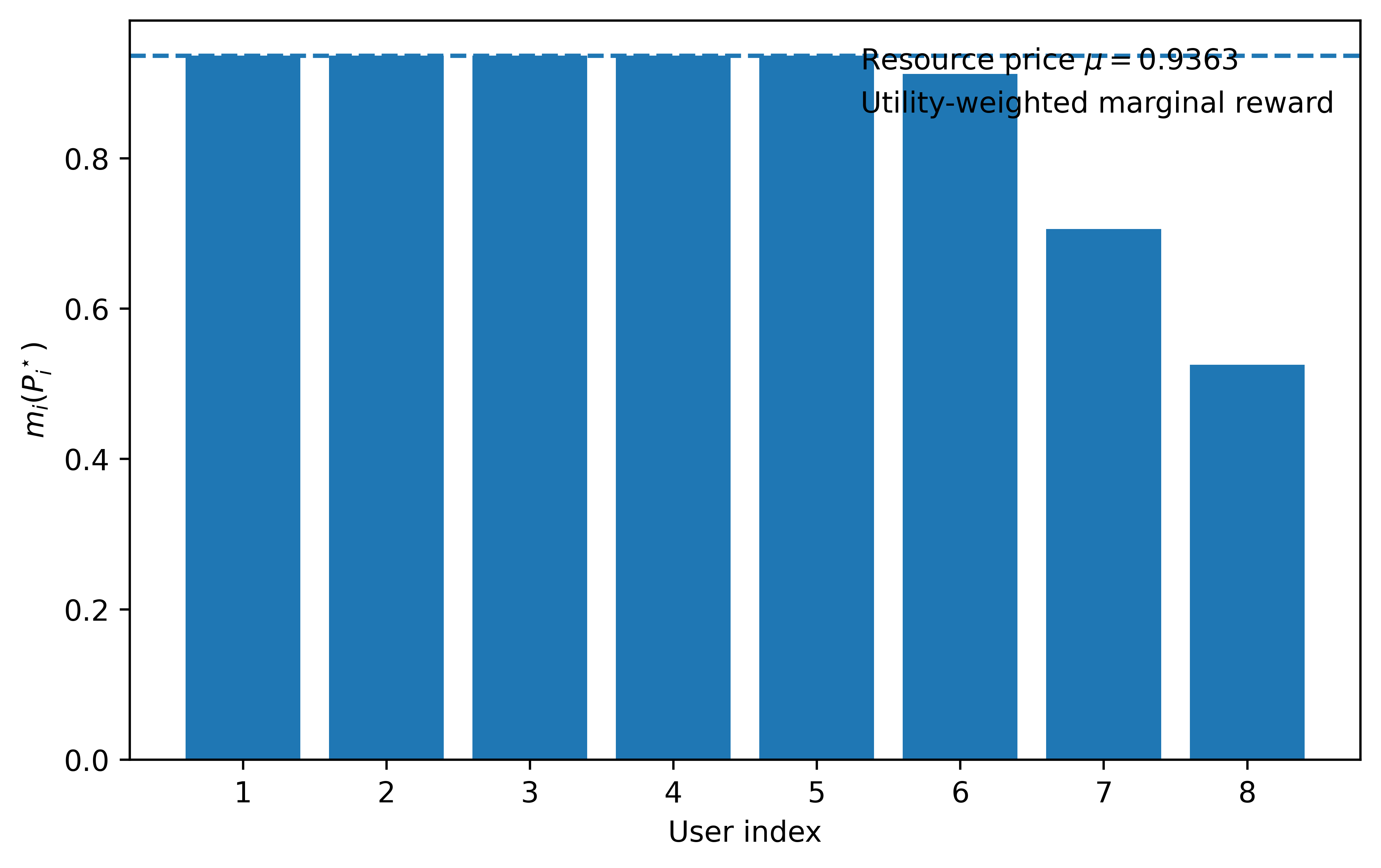}
    \caption{Utility-weighted marginal rewards at the globally optimal orthogonal allocation for \eqref{eq:wf_validation_utility}. Active users equalize the resource price $\mu$, while inactive users remain below it.}
    \label{fig:water_filling_validation}
\end{figure}

At the computed optimum, since $w_i=1$, we evaluate $m_i(P_i^\star)=u_i'(R_i(P_i^\star))R_i'(P_i^\star)$. Five users are active. Their marginal rewards equalize at $\mu=0.9362615$ with $\max_{i:P_i^\star>0}|m_i(P_i^\star)-\mu|
=2.34\times10^{-8}$. For inactive users, the KKT condition $m_i(0)\leq\mu$ is satisfied with zero maximum violation. Therefore, Figure~\ref{fig:water_filling_validation} provides a numerical check of the generalized water-filling condition at the unique global optimum of this concave instance.

\subsection{Scenario-Based EUT/CPT Allocation}
We finally use a fully specified fixed finite scenario set with $N=8$, $K=300$, $P_{\rm tot}=8$, and $N_0=1$. The scenario probabilities are equiprobable, $p_k=1/K$. The channel power gains are generated once as independent lognormal samples
\begin{equation}\label{eq:stochastic_lognormal_model}
g_i^{(k)}\sim\operatorname{Lognormal}(\mu_i^{\ln},(\sigma_i^{\ln})^2),
\end{equation}
where the prescribed means and coefficients of variation are $\bar{\mathbf g}=[2.8,2.4,2.0,1.7,1.3,1.1,0.9,0.75]$, $\mathbf c=[0.5,0.5,0.5,0.5,0.2,0.2,0.2,0.2]$, and
\begin{equation}
\sigma_i^{\ln}=\sqrt{\ln(1+c_i^2)},\qquad
\mu_i^{\ln}=\ln\bar g_i-\frac{(\sigma_i^{\ln})^2}{2}.
\end{equation}
Note that independence is assumed for this numerical benchmark, but the general model in Section~\ref{sec:stochastic} does not require it. 

The deterministic baselines for WSR, PF and reference-dependent are obtained
by optimizing their respective deterministic objectives using the
mean channel gains of the sample $\bar g_i=K^{-1}\sum_{k=1}^{K} g_i^{(k)}$, $i=1,\ldots,N$, and are then evaluated under the same stochastic scenarios $K$.

The user reference rates are the same as in Table~\ref{tab:deterministic_baselines}, and the user utility is
\begin{equation}\label{eq:stochastic_user_utility}
u_i(r)=r-r_i^0-1.5\,(r_i^0-r)_+.
\end{equation}
The utility collected at the scenario-level is
\begin{equation}\label{eq:stochastic_collection}
G_k(\vP)=\frac{1}{N}\sum_{i=1}^{N}u_i\!\left(\log_2(1+g_i^{(k)}P_i)\right).
\end{equation}
At the orchestrator, we use the reference point $z_0=-0.15$ and the piecewise-linear Bernoulli utility
\begin{equation}\label{eq:stochastic_orch_utility}
U_{\rm orch}(z)=
\begin{cases}
z-z_0, & z\geq z_0,\\
2(z-z_0), & z<z_0.
\end{cases}
\end{equation}
The CPT decision weights use the Prelec function in \eqref{eq:Prelec_PWF} with $\theta=0.65$ and $\gamma=0.90$ for both gains and losses.

We compare the decision criteria on exactly the same fixed scenario set
and scenario probabilities. With the nondecreasing concave user utilities in
\eqref{eq:stochastic_user_utility} and the nondecreasing concave orchestrator utility in \eqref{eq:stochastic_orch_utility}, each scenario outcome $G_k(\vP)$ is concave in $\vP$, and therefore the finite-scenario EUT objective is concave
for this orthogonal benchmark. CPT remains rank-dependent and piecewise smooth. To avoid conflating criterion differences with the local SCA trajectory, the fixed-scenario EUT and CPT objectives are solved here by multistart constrained SQP: one equal-power initialization and nine Dirichlet simplex initializations, with a maximum of $2000$ iterations and objective tolerance $10^{-11}$. The best converged feasible solution for each criterion is reported. 

\begin{table}[t]
\caption{Fixed-scenario EUT/CPT comparison. The values are empirical statistics for the single $K=300$ scenario set generated by \eqref{eq:stochastic_lognormal_model}, not Monte Carlo averages over regenerated scenario sets.}
\label{tab:stochastic_eut_cpt}
\centering
\begin{tabular}{lcccc}
\hline
Policy & EUT & CPT & $\overline{G}$ & $5\%$ $G$ \\
\hline
Equal power & 0.219 & 0.183 & 0.081 & -0.249 \\
WSR-det. & 0.218 & 0.185 & 0.078 & -0.214 \\
PF-det. & 0.176 & 0.138 & 0.045 & -0.295 \\
Ref. dep. utility & 0.255 & 0.226 & 0.112 & -0.185 \\
EUT-opt. & 0.258 & 0.227 & 0.115 & -0.179 \\
CPT-opt. & 0.257 & 0.228 & 0.114 & -0.182 \\
\hline
\end{tabular}
\end{table}

\begin{figure}[t]
    \centering
    \includegraphics[width=0.75\linewidth]{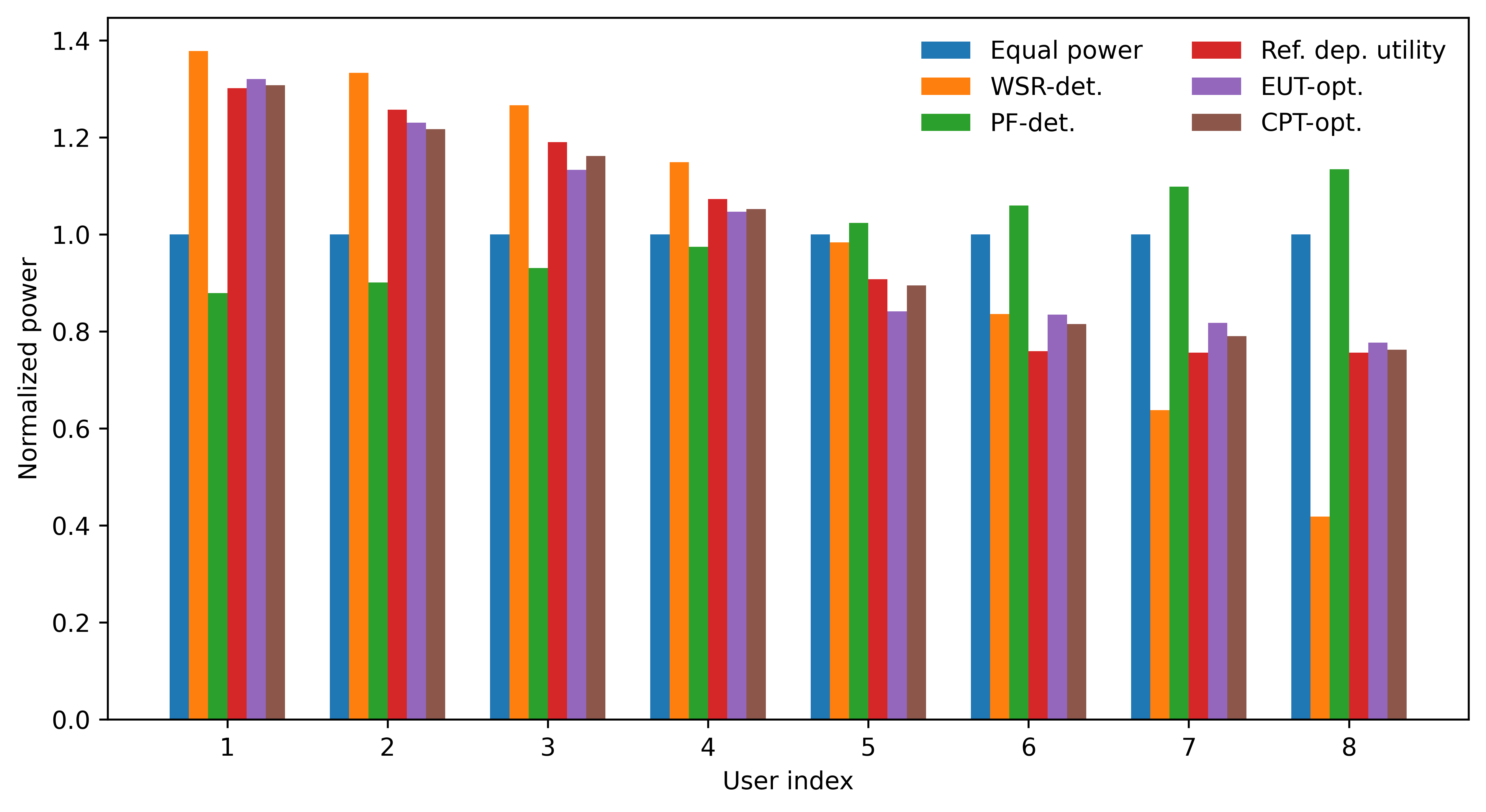}
    \caption{Normalized power allocations for the fully specified fixed-scenario policies in Table~\ref{tab:stochastic_eut_cpt}.}
    \label{fig:stochastic_allocations}
\end{figure}

For this fixed scenario set, the EUT-optimized policy attains the largest
reported EUT value, while the best CPT solution found by the multistart
SQP procedure attains the largest reported CPT value among the evaluated
policies. The two optimized allocations are close but not identical: relative to EUT optimization, CPT optimization slightly increases the CPT criterion while slightly reducing the EUT criterion and the raw mean $\overline G$. The ``5\% $G$'' column reports the empirical 5th percentile of the scenario outcomes $\{G_k(\vP)\}_{k=1}^{K}$. The 5\% $G$ column is reported separately because the CPT value is neither a percentile nor a CVaR measure.

\section{Conclusion}
This paper developed a generalized resource allocation framework that separates operational rewards, user-side utilities, and system-level decision criteria. Motivated by semantic, goal-oriented, and service-aware networks, the formulation captures how the value of an operational metric depends on the user objective, service reference point, and operating context. Reference-dependent utilities model heterogeneous service/QoS reference
levels, deficit sensitivity, post-reference saturation, and loss-aversion-like behavior.

The framework recovers weighted sum-reward maximization,
weighted proportional fairness, weighted $\alpha$-fairness, and stochastic expected-utility maximization as exact special cases, while soft QoS-aware allocation is obtained through a deficit-penalty generalization. The soft-penalty analysis gives an explicit $O(1/\lambda)$ excess-deficit bound, which becomes an
$O(1/\lambda)$ aggregate soft-QoS deficit-measure bound
when the QoS targets of the positively weighted users are jointly feasible, under global solution of the penalized problem. 

For deterministic orthogonal systems, utility-weighted marginal-reward equalization is a necessary KKT condition at differentiable local optima and is sufficient for global optimality when each composite user objective is concave. Strict concavity yields a unique generalized water-filling response. For Shannon rewards, the pointwise condition $u_i''(r)\leq(\ln2)u_i'(r)$ provides an explicit
composite-concavity test and shows that limited convexity of the deficit branch is compatible with global concavity. 

For nonorthogonal and stochastic systems, the smooth SCA framework yields first-order stationary accumulation points under the stated assumptions, while the composite nonsmooth formulation yields Clarke-stationary accumulation points. For CPT-based optimization, the corresponding conclusion is branch-stationarity under the local nondegenerate ordering/reference-point assumptions stated in the analysis. The smooth fixed-step specialization additionally admits an $O(1/\epsilon^2)$ best-response-residual bound.

Numerical results illustrate how utility curvature, reference levels, deficit sensitivity, channel quality, and resource budgets alter marginal resource values and the resulting allocation. The fixed-scenario experiment further shows that EUT and CPT can select different policies on the same scenario set and scenario probabilities, without implying out-of-sample superiority of either criterion. These results motivate accounting jointly for operational efficiency and user-side value in semantic and goal-oriented communication systems. Future work includes distributed
implementations, online utility learning, stronger convergence analysis across CPT ordering changes, and extensions to queue-aware, latency-sensitive, and large-scale stochastic networks.

\section*{Acknowledgments}
This work was supported in part by the European Research Council (ERC) under the EU's Horizon 2020 research and innovation programme (Grant agreement No. 101003431), which partially supports the work of M. Kountouris. The work of S. Vaidanis and P. Stavrou was supported by the SNS JU project 6G-GOALS \cite{strinati:2024} under the EU's Horizon programme Grant Agreement No. 101139232. S. Vaidanis was also supported by the Onassis Foundation - Scholarship ID: F ZU 076-1/2024-2025.

\bibliographystyle{IEEEtran}
\bibliography{my_bibliography}

\end{document}